\documentclass{article}
\usepackage[top=3cm , bottom=2.5cm , left=2.5cm , right=2.5cm]{geometry}
\usepackage{booktabs}
\usepackage[justification=justified]{caption}
\usepackage{subcaption}
\usepackage{amsfonts} 
\usepackage{abstract}
 \usepackage{bibentry}
\usepackage{float}
\usepackage{amsmath,amsthm,mathtools,amsfonts,bm}
\usepackage{caption}
\usepackage{subcaption}
\usepackage{multirow}
\usepackage[stable]{footmisc}
\usepackage{abstract}
\usepackage[table]{xcolor}
\usepackage[titletoc]{appendix}
 \usepackage{bibentry}
\usepackage[bookmarks=false]{hyperref}
\usepackage{authblk}

\hypersetup{
    colorlinks=true,
allcolors=blue,
}

\newcommand{\ra}[1]{\renewcommand{\arraystretch}{#1}}
\title{Non-reciprocity using quadrature-phase time-varying slab resonators}
\author[1]{Mahdi Chegnizadeh}
\author[1*]{Mohammad Memarian}
\author[1]{Khashayar Mehrany}

\affil[1]{Department of Electrical Engineering, Sharif University of Technology, Tehran, Iran}

\affil[*]{Corresponding author: mmemarian@sharif.ir}

\begin{document}

\maketitle
\begin{abstract}
In this paper, it is shown that non-reciprocity can be observed in time-varying media without employing spatio-temporal modulated permittivities. We show that by using only two one dimensional Fabry-Perot slabs with time-periodic permittivities having quadrature phase difference, it is possible to achieve considerable non-reciprocity in transmission at the incidence frequency. To analyze such scenario,generalized transfer matrices are introduced to find the wave amplitudes of all harmonics in all space. The results are verified by in-house FDTD simulations. Moreover, in order to have a simple model of such time-varying slab resonators, a time-perturbed coupled mode theory is developed for multiple resonances, and it is shown that the results obtained by this method and the analytical method are in excellent agreement.. 
\end{abstract}\section{Introduction}
Electromagnetic wave propagation in time-varying media has gained considerable research attention in recent years, owing to the promises it delivers for various applications. A chief promise is achieving non-reciprocity without using magnetic materials. At microwave frequencies, such non-reciprocity may yield alternative solutions to existing components such as isolators \cite{wang2014time,wang2012non,taravati2017nonreciprocal,nagulu2018nonreciprocal}, gyrators, and circulators \cite{nagulu2018nonreciprocal}. Non-reciprocity in silicon photonics using time-varying methods was first proposed theoretically in \cite{yu2009complete} and was implemented in a silicon chip \cite{lira2012electrically}. Other applications such as frequency combs \cite{dong2008inducing,preble2007changing,tanabe2009dynamic} and time-varying metasurfaces \cite{chamanara2019simultaneous,hadad2015space,chamanara2016spacetime,shaltout2015time,taravati2017nonreciprocal2} are also introduced by using time-varying refractive indices. 

In microwave transmission lines, non-reciprocity can be achieved by using a line capacitance which is modulated with both space and time dependency \cite{wang2014time,wang2012non} and this modulation has a propagating wave function, i.e. $C(z,t)=C(\Omega t-\beta z)$. Similarly, in order to achieve non-reciprocity in a bulk medium, it is common to use a permittivity function with both spatial and temporal modulation \cite{taravati2017nonreciprocal,chamanara2017new,taravati2019generalized,taravati2019space}. In such work, it is typically assumed that such spatio-temporal modulated permittivities can be achieved by sending a travelling acoustic wave through the medium, although this is hard to actually implement in practice due to the partial reflections of the acoustic wave from boundaries. In silicon chips \cite{lira2012electrically}, the spatio-temporal modulation of the refractive index is achieved by discrete changes in space. A primary goal of this paper is to propose a method to achieve non-reciprocity for wave propagation in a 1D medium without modulating permittivities in space, but only in time. 

In this work, we show that two 1D Fabry Perot slab resonators, which are only time modulated but are uniform spatially, can indeed result in significant non-reciprocity. This point is demonstrated by using a general solution based on transfer matrices \cite{anemogiannis1999determination} for finding the wave amplitudes in each FP slab. This method is based on Floquet-Bloch theorem and field matching at boundaries. To have a better understanding of the underlying physics of time-varying FP resonators, a simple model for a medium consisting of one or more time-varying FP slabs is proposed based on temporal coupled mode theory (TCMT) \cite{fan2003temporal}. We phenomenologically perturb the TCMT of time-invariant resonators to find the behavior of resonators with time-varying permittivities (for both single mode and multi-mode resonators). It will be shown that the results of both the analytical solution (transfer matrices) and the perturbed TCMT method virtually overlap. Finally, it is demonstrated that by using only two same-sized FP slabs placed at a specific distance from each other whose permittivities are sinusoidally modulated in time with a quadrature phase difference, a noticeable level of non-reciprocity can be achieved at the incidence frequency. It should be noted that achieving non-reciprocity in the harmonics of a time-modulated structure might be easier, but at the incidence frequency (zeroth harmonic) is a much challenging and desirable result. Such interesting results are validated by using in-house FDTD simulations. 


\section{Single Time-Varying Fabry-Perot Slab}
In this section, the transmission of normal incident wave on a single FP slab whose permittivity changes periodically with time is analyzed. This problem has already been solved in \cite{zurita2009reflection}, where Floquet-Bloch theorem and field matching at boundaries was employed to solve the problem. However, the solution was limited to a single slab, and no general solution was suggested that could handle multiple slabs or layers with different permittivity values. 

\begin{figure}[h!]
\centering\includegraphics[width=8cm]{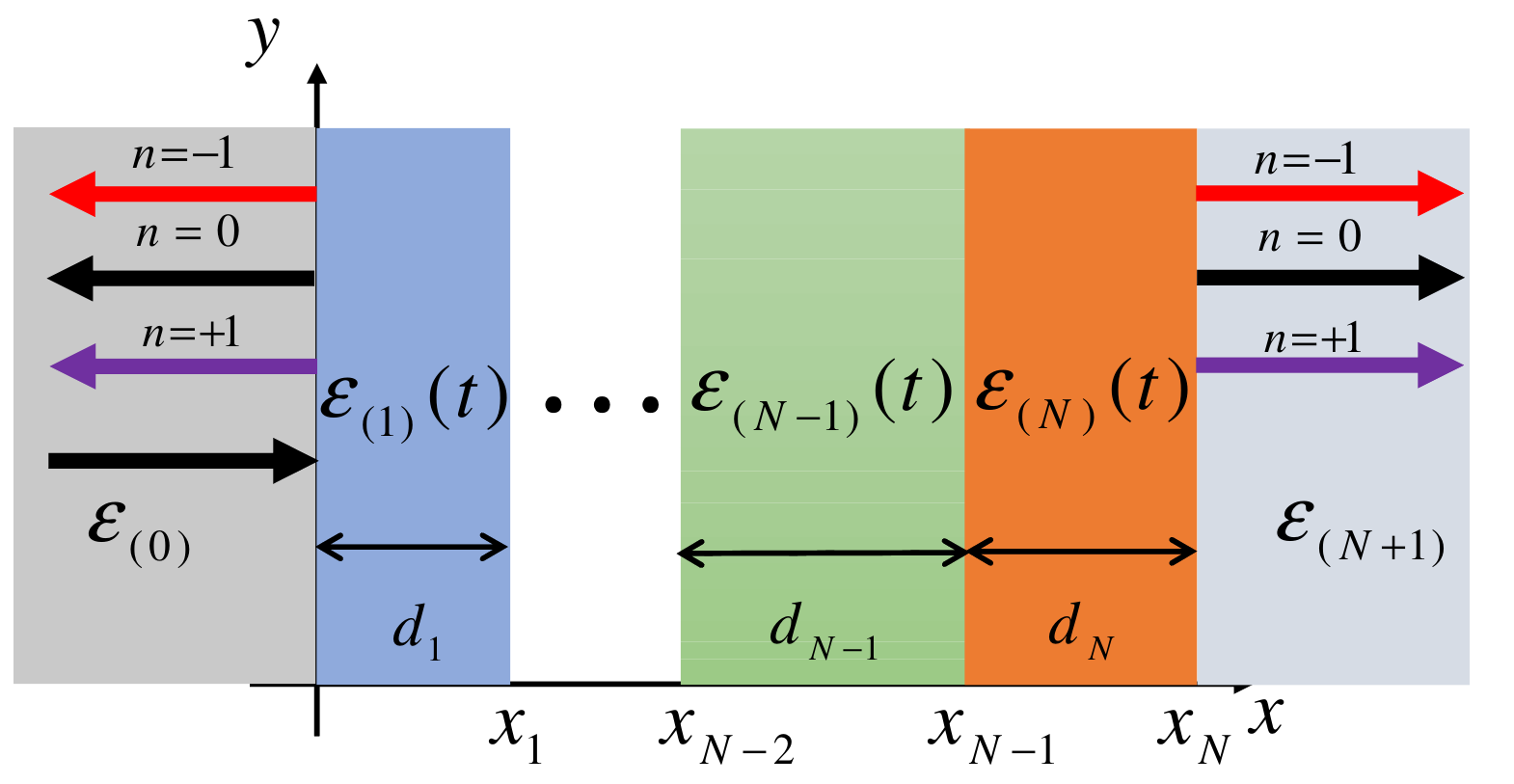}
\caption{Normal incidence on multiple time varying slabs.}
\label{fig:structure}
\end{figure}

In Appendix A, we outline a generalized transfer matrix method that can yield the wave amplitudes of all generated harmonics allover the space for normal wave incident on multiple slabs whose permittivities changes periodically in time with frequency $\Omega$ (Fig. (\ref{fig:structure})). By considering a forward incident wave in the first layer at frequency $\omega$ and using Floquet-Bloch theorem, we can write the field in the incidence layer as ${E_z}(x,t) = {e^{j\left( {\omega t - {\beta _0}x} \right)}} + \sum\limits_{n =  - N}^{ + N} {{R_n}{e^{j\left( {(\omega  + n\Omega )t + {\beta _n}x} \right)}}}  + c.c.$ and in the last layer as ${E_z}(x,t) = \sum\limits_{n =  - N}^{ + N} {{T_n}{e^{j\left( {(\omega  + n\Omega )t - {\alpha _n}x} \right)}}}  + c.c.$, where $R_n$ and $T_n$ are unknown reflection and transmission coefficients of generated harmonics that can be determined by using the following equation:
\begin{equation}
\left[ {\begin{array}{*{20}{c}}
{{{\bf{T}}_{(2N + 1) \times 1}}}\\
{{{\bf{0}}_{(2N + 1) \times 1}}}
\end{array}} \right] = \left\{ {\prod\limits_{i = 0}^M {{{\bf{Q}}_{(M - i) \to (M + 1 - i)}}} } \right\}\left[ {\begin{array}{*{20}{c}}
{{{\rm{I}}_{(2N + 1) \times 1}}}\\
{{{\bf{R}}_{(2N + 1) \times 1}}}
\end{array}} \right]
\label{eq:1000}
\end{equation}
where $\bf T$, $\bf R$, and $\bf I$ are the column vectors of transmission coefficients, reflection coefficients, and incident wave, respectively, and ${\bf Q}_{i\to i+1}$ is the transfer matrix that relates the field amplitudes of $i$th layer to $(i+1)$th layer. The elements of the transfer matrices are introduced in more detail in Appendix A. The interested reader can investigate that the proposed formulation can accurately reproduce the results of \cite{zurita2009reflection}; however, we do not bring the results here for the sake of brevity. 
\begin{figure}
\centering
\begin{subfigure}[b]{0.4\textwidth}
   \includegraphics[width=1\linewidth]{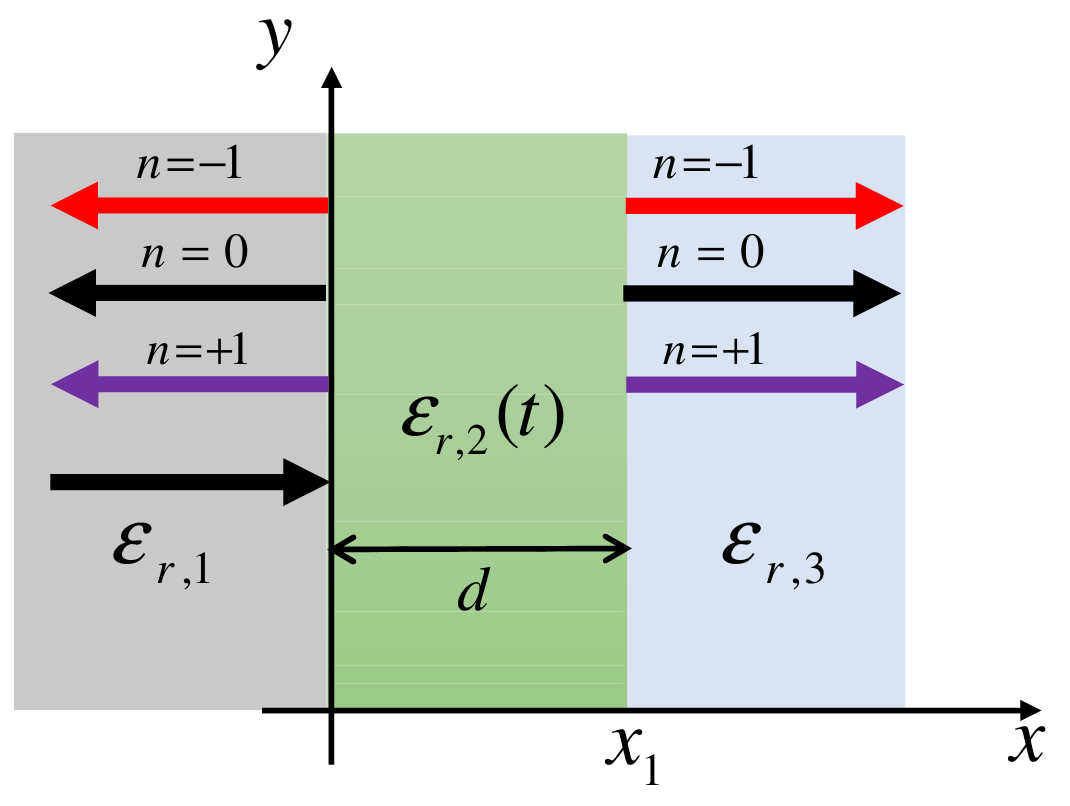}
   \caption{}
   \label{fig:one_res} 
\end{subfigure}

\begin{subfigure}[b]{0.4\textwidth}
   \includegraphics[width=1\linewidth]{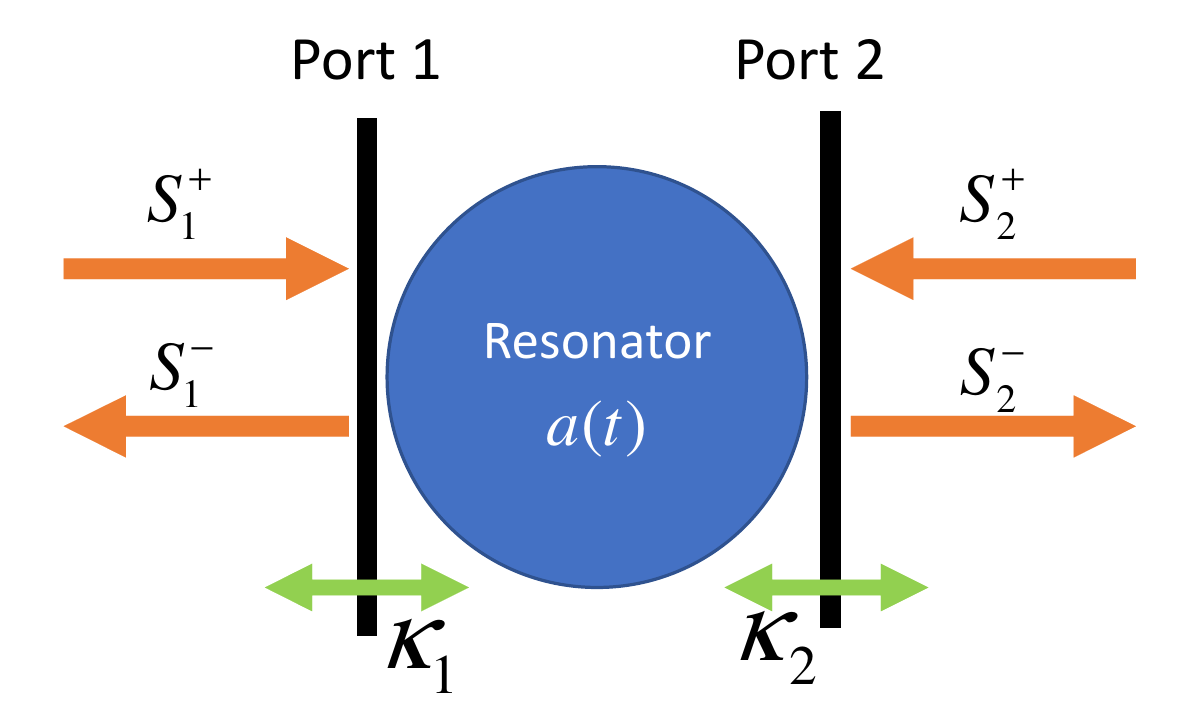}
   \caption{}
   \label{fig:resonators}
\end{subfigure}
\label{fig:1}
\caption{(a) Normal incidence on a time varying slab and (b) the model used for TCMT.}
\end{figure}

The formulation presented provides accurate results in terms of the transmitted and reflected waves and harmonics of a time-varying FP slab (Fig. \ref{fig:one_res}). However, it does not provide much insight in to the physics of the structure and the resonance phenomenon taking place. Thus, we try to model the behavior of the time-varying (TV) slab using TCMT (Fig. \ref{fig:resonators}). Here, we consider a sinusoidal time-varying permittivity, i.e. $\varepsilon_{r,2}(t)=\varepsilon_{r,2}+\Delta \varepsilon_2 \cos(\Omega t)$. To this end, first the transmission and reflection from a FP slab with $\Delta \varepsilon_2=0$ (time-invariant permittivity) is found using TCMT. Subsequently, the formulation is perturbed phenomenologically to account for the transmission and reflection from a slab with $\Delta \varepsilon_2\neq0$ (TV slab). In both mentioned cases, we assume that modulation frequency ($\Omega$) is known.

A FP resonator has infinite number of resonance modes. However, in the vicinity of a specific resonance, one can model it with the resonance of a single mode resonator. The TCMT relations for a single mode resonator can be written as \cite{fan2003temporal}:
\begin{subequations}
\begin{equation}
\frac{{da}}{{dt}} = \left( {j{\omega _0} - \frac{1}{\tau }} \right)a + ({\left\langle \kappa  \right|^*})\left| {{S_ + }} \right\rangle 
\label{eq:15a}
\end{equation}
\begin{equation}
\left| {{S_ - }} \right\rangle  = {\bf{C}}\left| {{S_ + }} \right\rangle  + a\left| \kappa  \right\rangle 
\label{eq:15b}
\end{equation}
\label{eq:15}
\end{subequations}
where $\omega_0$ is the resonance frequency, $\tau$ is the lifetime of the resonance, $\left| \kappa  \right\rangle$ is the vector of the coupling factors exchanging energy between resonator and outside, ${\bf C}$ is the matrix that couples the input power from the ports directly to the output ports, and $|a|^2$ is proportional to the energy of the resonator. The relations of these parameters in more details can be found in \cite{fan2003temporal}. 

Here, due to similarity between the FP resonator and a direct coupled resonator, we consider that the matrix $\bf C$ is an identity matrix. This resonator has two physical ports; hence, all column vectors are $2\times 1$, and all matrices are $2 \times 2$. One can easily show that $S_2^-$ (transmission) can be calculated as:
\begin{equation}
\frac{S_2^-}{S_1^+}  =  \pm \frac{{|{\kappa _1}|\sqrt {2/\tau  - |{\kappa _1}{|^2}} }}{{j(\omega  - {\omega _0}) + 1/\tau }}
\label{eq:16}
\end{equation}
where we have used the fact that $|{\kappa _1}|^2+|{\kappa _2}|^2=2/\tau$, and the $\pm$ is related to the parity of the mode. For a case of $\varepsilon_{r,1}=\varepsilon_{r,3}=1$ in Fig. \ref{fig:one_res} (media on right and left of the slab being the same) then $|{\kappa _1}|=|{\kappa _2}|=\sqrt{1/\tau}$. Thus the two unknown coefficients of Eq. (\ref{eq:16}) ($\tau$ and $\omega_0$) can be found from the result of analytical solution presented earlier. The resonance frequency ($\omega_0$) is the frequency at which transmission reaches its maximum (in this case becomes 1), and $\tau$ can be found from the full width at half maximum of the transmission frequency response.

\begin{figure}[h!]
\centering\includegraphics[width=10cm]{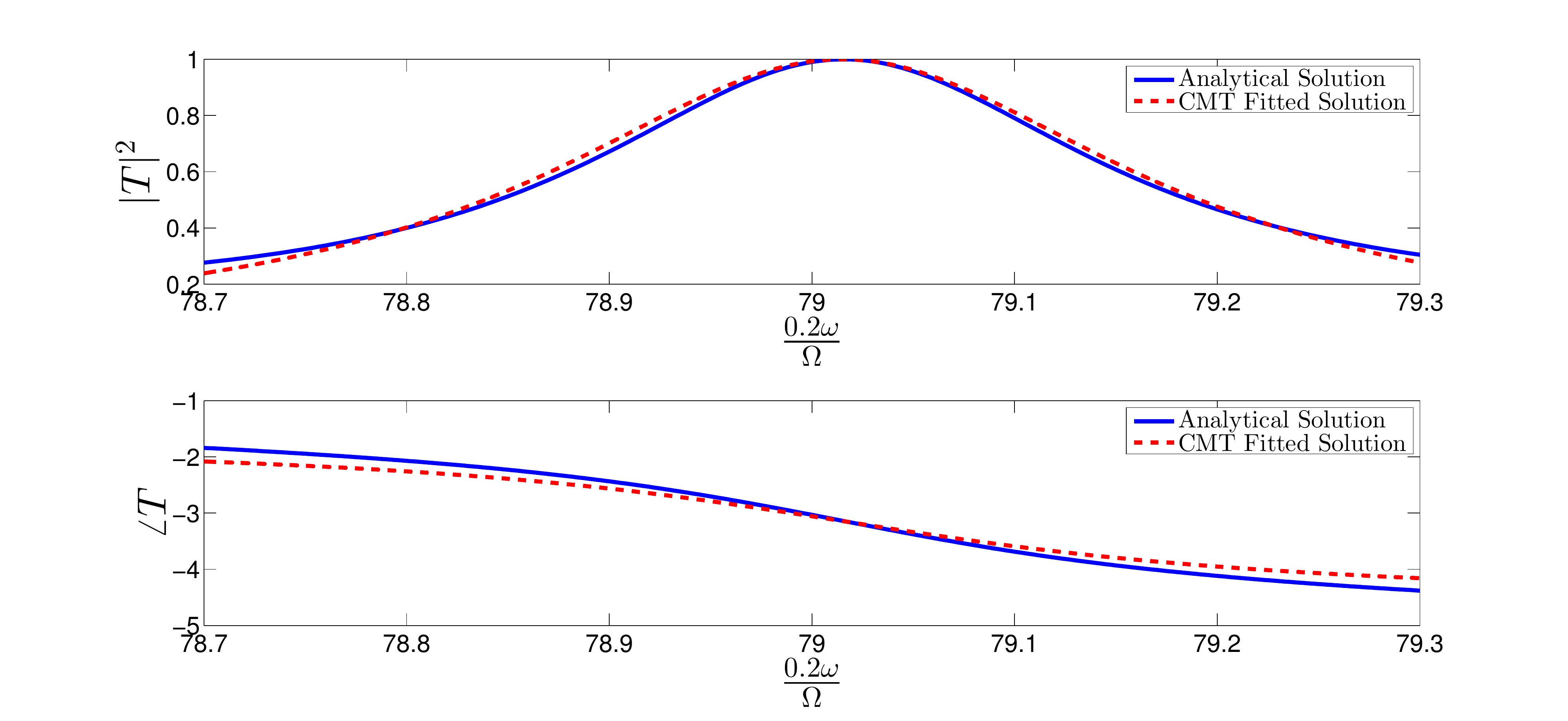}
\caption{TCMT fitted solution for a single time-invariant slab.}
\label{fig:Fitted_oneslab_noMod}
\end{figure}

For example, for a FP slab with $\frac{\Omega}{c_0}d = 3.3\times 0.2/\sqrt{16}$ whose background permittivity is $\varepsilon_r=16$, the analytical solution and the fitted one by using Eq. (\ref{eq:16}) can be seen in Fig. \ref{fig:Fitted_oneslab_noMod}. This figure is plotted  versus incidence frequency divided by $\Omega$ to make it dimensionless. As it is apparent from this figure, both solutions virtually overlap, which means that the proposed solution using TCMT is is accurately modeling the resonance behavior. 

Having found the TCMT model for the static resonator, we then need to find the transmission of wave through the time-varying FP slab. This is achieved by adding perturbation terms to Eq. (\ref{eq:15}). We postulate that by using a time-varying permittivity in the slab resonator, its resonance frequency and the coupling coefficients will change (ever so slightly) due to the small changes in permittivity. Because the perturbation varies periodically with time (with frequency $\Omega$), a periodic perturbation (with three terms) is considered as follows for both the complex resonance frequency ($\alpha_n$ terms) and the coupling coefficients ($\chi_n$ terms):
\begin{equation}
\frac{{da}}{{dt}} = \left( {j{\omega _0} - \frac{1}{\tau } + \sum\limits_{n =  - 1}^{ + 1} {{\alpha _n}{e^{jn\Omega t}}} } \right)a + \left( {{\kappa _1} + \sum\limits_{n =  - 1}^{ + 1} {{\chi _n^{(1)}}{e^{jn\Omega t}}} } \right)S_1^ + 
\label{eq:17}
\end{equation}
For a wave incident on port 1 at frequency $\omega$, we can consider $S_1^+=e^{j\omega t}$. As a result, the solution of this equation should have the following form: 
\begin{equation}
a = \sum\limits_{n =  - 1}^{ + 1} {{a_n}{e^{j(\omega  + n\Omega )t}}} 
\label{eq:18}
\end{equation}
Equation (\ref{eq:17}) may therefore be written in phasor form as:
\begin{equation}
\left( {{{\bf{W}}_0} - \Delta {\bf{W}}} \right)\left( {\left| {{{\bf{a}}_0}} \right\rangle  + \left| {\delta {\bf{a}}} \right\rangle } \right) = \left| {{{\mathbf{\kappa }}_0}^{(1)}} \right\rangle  + \left| {\delta {\mathbf{\kappa }^{(1)}}} \right\rangle 
\label{eq:19}
\end{equation}
where
\begin{equation}
\begin{aligned}
\left| {{{\bf{\kappa }}_0}^{(1)}} \right\rangle  &= {\left[ {\begin{array}{*{20}{c}}
0&{{\kappa _1}}&0
\end{array}} \right]^T}\\
\left| {\delta {\bf{\kappa }}^{(1)}} \right\rangle & = {\left[ {\begin{array}{*{20}{c}}
{{\chi _{ + 1}^{(1)}}}&{{\chi _0}^{(1)}}&{{\chi _{ - 1}^{(1)}}}
\end{array}} \right]^T}\\
{{\bf{W}}_0}_{,(n,m)} &= \left\{ {j(\omega  - {\omega _0} + (2 - n)\Omega ) + 1/\tau } \right\}{\delta _{nm}}\\
\Delta {{\bf{W}}_{(n,m)}} &= {\alpha _{m - n}}
\end{aligned}
\label{eq:20}
\end{equation}
where ($\left| {{{\bf{a}}_0}} \right\rangle  + \left| {\delta {\bf{a}}} \right\rangle $) is now a vector containing energy amplitudes of the different time-harmonics of the resonator. By writing the solution of the Eq. (\ref{eq:17}) in the form of Eq. (\ref{eq:19}), we have also divided it into two parts: the unperturbed ($ \left| {{{\bf{a}}_0}} \right\rangle$) and the perturbation ($\left| {\delta {\bf{a}}} \right\rangle$). The main reason for doing this is to avoid inverting matrices such as $\Delta \bf W$ which are approximately full. The unperturbed solution is:
\begin{equation}
\left| {{{\bf{a}}_0}} \right\rangle  = {\bf{W}}_0^{ - 1}\left| {{{\bf{\kappa }}_0}^{(1)}} \right\rangle  = {\left[ {\begin{array}{*{20}{c}}
0&{\frac{{{\kappa _1}}}{{j(\omega  - {\omega _0}) + 1/\tau }}}&0
\end{array}} \right]^T}
\label{eq:21}
\end{equation}
Equation (5) can also be further approximated as $\left| {\delta {\bf{a}}} \right\rangle \simeq {\bf W}_0^{-1}\Delta {\bf W}  \left| {{{\bf{a}}_0}}\right\rangle+{\bf W}_0^{-1}\left| {\delta {\mathbf{\kappa }}} \right\rangle $, where second order perturbation terms are discarded (i.e. assuming $ \Delta {\bf{W}} \left| {\delta {\bf{a}}} \right\rangle \simeq  \mathbf{0}$)

The perturbed form of transmission can then be written as:
\begin{equation}
\frac{S_2^ -}{S_1^ +}  = \left( {{\kappa _2} + \sum\limits_{n =  - 1}^{ + 1} {\chi _{_n}^{(2)}{e^{jn\Omega t}}} } \right)\sum\limits_{n =  - 1}^{ + 1} {{a_n}{e^{j(\omega  + n\Omega )t}}} 
\label{eq:22}
\end{equation}
By using this equation, transmission at the incident and ($\pm1$) harmonics ($\omega\pm\Omega$) can be written as:
\begin{subequations}
\begin{equation} 
\begin{aligned}
{\left. {\frac{{S_2^ - }}{{S_1^ + }}} \right|_\omega }  = \frac{{{(\kappa _1+\chi _{_0}^{(1)})}{(\kappa _2+\chi _{_0}^{(2)})}}}{{j(\omega  - {\omega _0}) + 1/\tau }} + \frac{{\chi _{_{ - 1}}^{(2)}\chi _{_{ + 1}}^{(1)}}}{{j(\omega  - {\omega _0} + \Omega ) + 1/\tau }} + \frac{{\chi _{_{ - 1}}^{(1)}\chi _{_{ + 1}}^{(2)}}}{{j(\omega  - {\omega _0} - \Omega ) + 1/\tau }}
\end{aligned}
\label{eq:25a}
\end{equation}
\begin{equation}
{\left. {\frac{{S_2^ - }}{{S_1^ + }}} \right|_{\omega  + \Omega }} = \frac{{\chi _{_{+1}}^{( 2)}\left( {{\kappa _1} + \chi _{_0}^{(1)}} \right)}}{{j(\omega  - {\omega _0}) + 1/\tau }} + \frac{{\chi _{_{+1}}^{( 1)}\left( {{\kappa _2} + \chi _{_0}^{(2)}} \right)}}{{j(\omega  - {\omega _0} + \Omega ) + 1/\tau }}\label{eq:25b}
\end{equation}
\begin{equation}
{\left. {\frac{{S_2^ - }}{{S_1^ + }}} \right|_{\omega  - \Omega }} = \frac{{\chi _{_{-1}}^{(2)}\left( {{\kappa _1} + \chi _{_0}^{(1)}} \right)}}{{j(\omega  - {\omega _0}) + 1/\tau }} + \frac{{\chi _{_{-1}}^{(1)}\left( {{\kappa _2} + \chi _{_0}^{(2)}} \right)}}{{j(\omega  - {\omega _0} - \Omega ) + 1/\tau }}\label{eq:25c}
\end{equation}
\label{eq:25}
\end{subequations}
Two scenarios may be envisioned based on the symmetry of the structure, i.e. whether $\varepsilon_{r,1}=\varepsilon_{r,3}$ or not. In the following, we investigate the possibility of achieving non-reciprocity in both cases.

\subsection{Parity Symmetric Structure}
\begin{figure}[h!]
\centering\includegraphics[width=10cm]{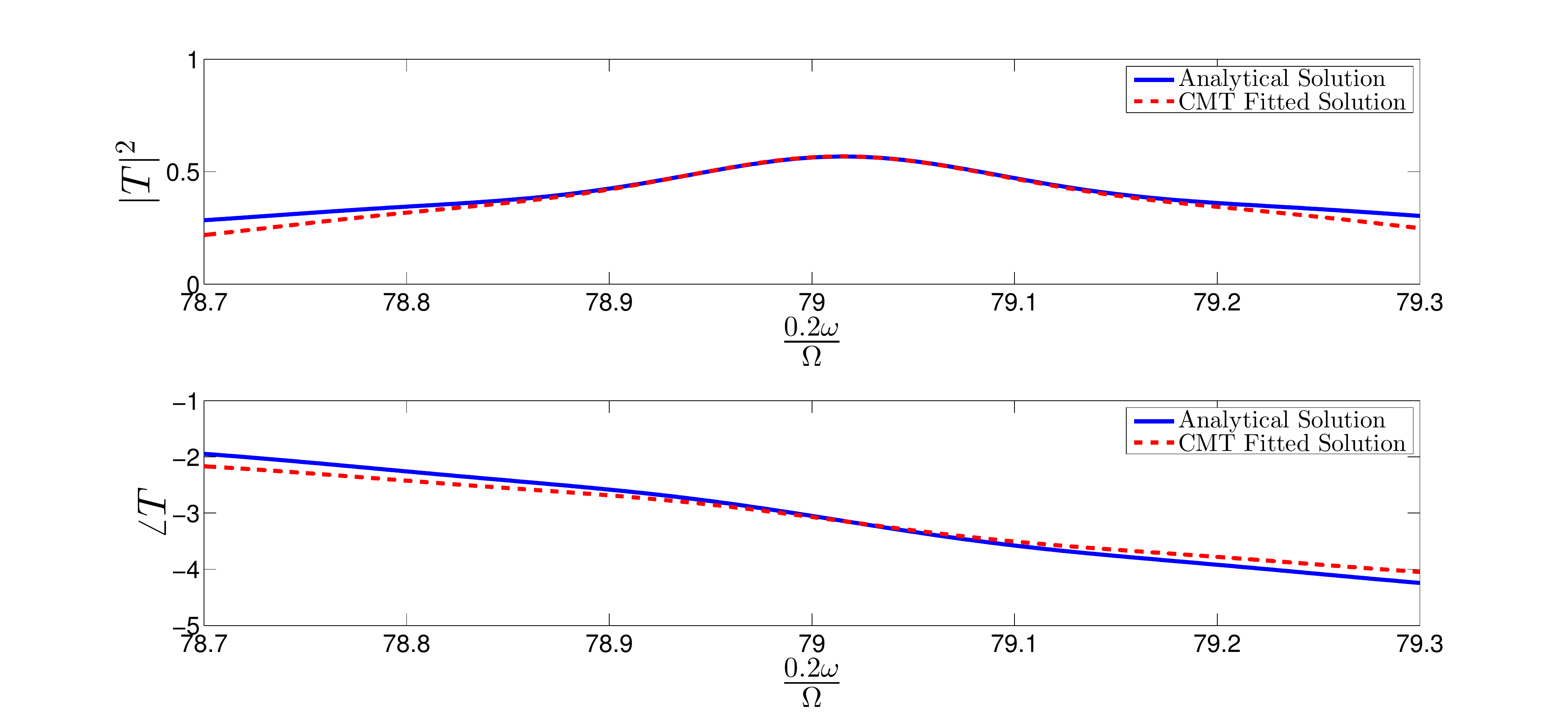}
\caption{TCMT fitted solution for transmission of incident frequency from a time-varying slab versus incidence frequency.}
\label{fig:Fitted_oneslab_Mod}
\end{figure}

\begin{figure}[h!]
\centering\includegraphics[width=10cm]{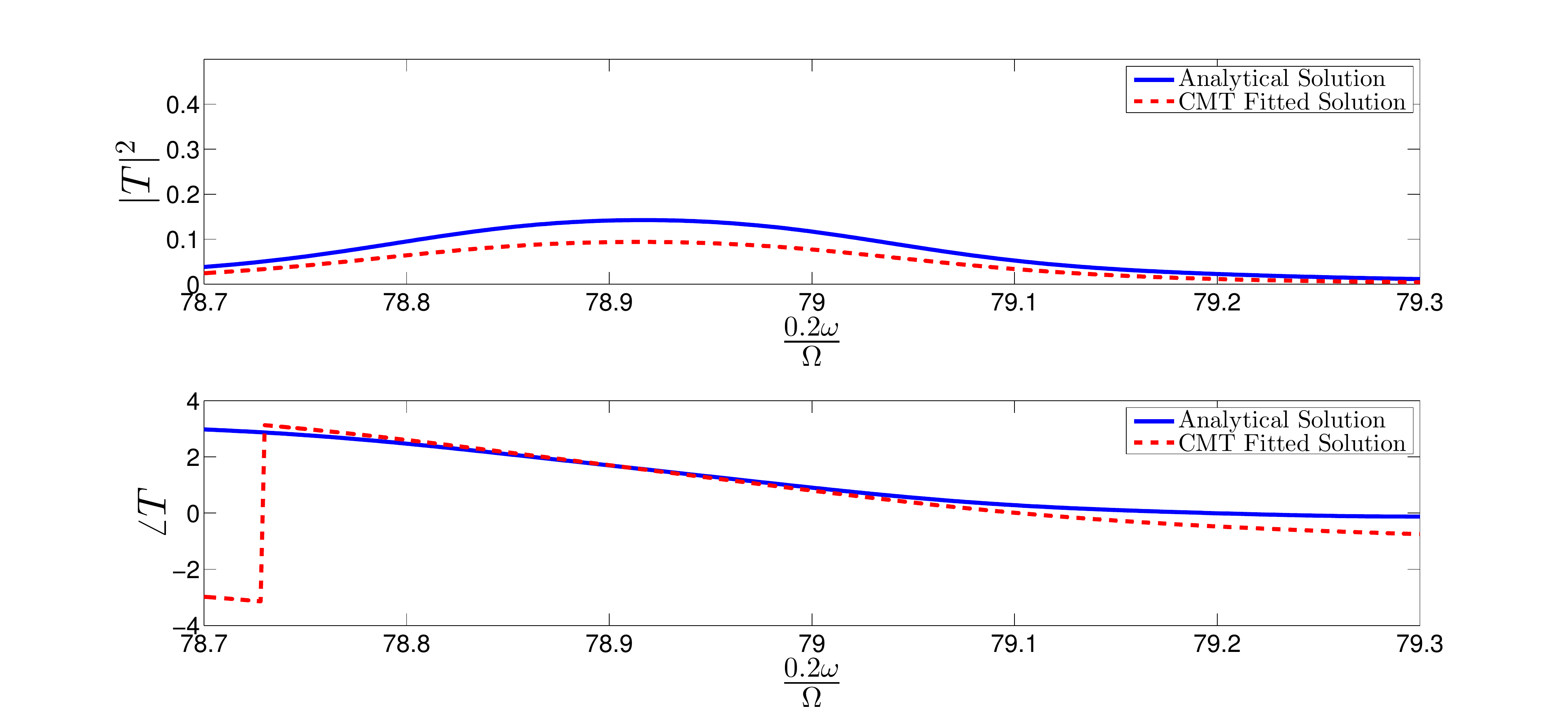}
\caption{TCMT fitted solution for transmission of $(+1)$ harmonic from a time-varying slab versus incidence frequency.}
\label{fig:Fitted_oneslab_Mod_1}
\end{figure}

For a parity symmetric scenario ($\varepsilon_{r,1}=\varepsilon_{r,3}$), the coupling factors to the two ports and their perturbations are equal, thus $\kappa_1=\kappa_2=\kappa=\pm j\sqrt{1/\tau}$ and $\chi_n^{(1)}=\chi_n^{(2)}=\chi_n$. It is evident from Eq. (\ref{eq:25}) and its similar equation for $S_1^- / S_2^+$, that they are equal and thus we always have reciprocity, as expected. Equation (\ref{eq:25}) will have two fitting terms ($\chi_0$ and $\chi_{+1}\chi_{-1}$), which can be easily found by fitting to the analytical results. For example, for a sinusoidal permittivity of $\varepsilon_{r,2}(t)=16+0.075 \cos(\Omega t)$, the analytical and fitted solutions are plotted in Fig. \ref{fig:Fitted_oneslab_Mod}. As it can be seen in this figure, both solutions virtually overlap. In addition, to show that this is a valid fitting, the transmission of the ($+1$) harmonic at $\omega+\Omega$ is also calculated from Eq. (\ref{eq:25b}) by using the fitting parameters that were obtained from fitting the zeroth harmonic (Eq. (\ref{eq:25a})). The result can be seen in Fig. \ref{fig:Fitted_oneslab_Mod_1}, where both the fitted solution and the analytical results overlap. 

\subsection{Parity Asymmetric Structure}
Although one might expect that non-reciprocity can be achieved by breaking the parity-symmetry, the numerical study of the problem shows that a single time varying FP slab resonator never shows non-reciprocity at the incident frequency, i.e. the zeroth time harmonic. Thanks to the TCMT, this point can be justified by using Eq. (\ref{eq:25a}). Reciprocity in transmission, which is tantamount to the equation remaining unchanged by substituting 1 for 2 and vise versa, enforces us to have $\frac{\chi _{_{ + 1}}^{(1)}}{\chi _{_{ - 1}}^{(1)}}=\frac{\chi _{_{ + 1}}^{(2)}}{\chi _{_{ - 1}}^{(2)}}$. The fact that this condition is always met is not unexpected because the ratio of red-shifted and blue-shifted power injected to the resonator does not change by reversing the parity of the structure.

It should be however noted that non-reciprocity can be observed for the $\pm1$ harmonics ($\omega\pm\Omega$) in the asymmetrical structure. This can be explained by using Eqs. (\ref{eq:25b}) and (\ref{eq:25c}). If this medium was reciprocal at $\pm1$ harmonics, we should have $\frac{\chi _{_{ + 1}}^{(1)}}{\chi _{_{ + 1}}^{(2)}}=\frac{\chi _{_{ - 1}}^{(1)}}{\chi _{_{ - 1}}^{(2)}}=\frac{\chi _{_{ 0}}^{(1)}+\kappa_1}{\chi _{_{0}}^{(2)}+\kappa_2}$. We argued that the first equality is accepted due to the reciprocity at the incidence frequency; however, the second equality cannot be true, because it means that the ratio of the perturbed $\kappa$s of the two ports is equal to the ratio of the $\pm1$ perturbation terms of the two ports, which is counterintuitive. 

\begin{figure}[h!]
\centering\includegraphics[width=10cm]{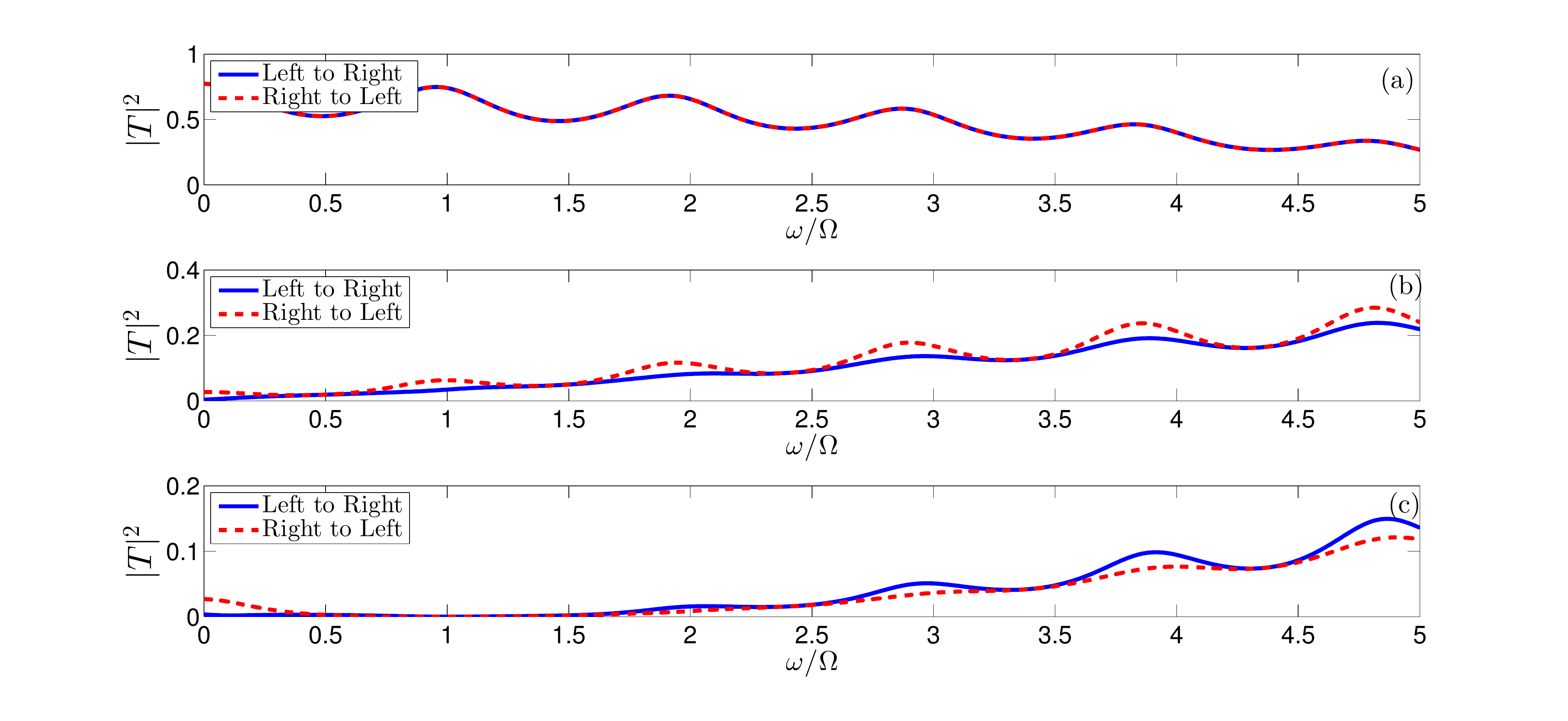}
\caption{Transmission from right to left (solid blue line) and left to right (dashed red line) at (a) incident frequency ($\omega$), (b) +1 harmonic ($\omega+\Omega$), and (c) -1 harmonic ($\omega-\Omega$).}
\label{fig:asmtrc}
\end{figure}

For example, for a FP slab with $\frac{\Omega}{c_0}d=3.3/\sqrt{16}$, we assign the permittivities as $\varepsilon_{r,1}=1$, $\varepsilon_{r,2}=16$, $\varepsilon_{r,3}=8$, and $\Delta\varepsilon_{2}=4$. The transmission at the zeroth, $+1$, and $-1$ harmonics can be seen in Fig. (\ref{fig:asmtrc}), where reciprocity at the incidence frequency and non-reciprocity at $\pm1$ harmonics are evident

\section{Two Cascaded Fabry-Perot Slabs}
\begin{figure}[h!]
\centering\includegraphics[width=6cm]{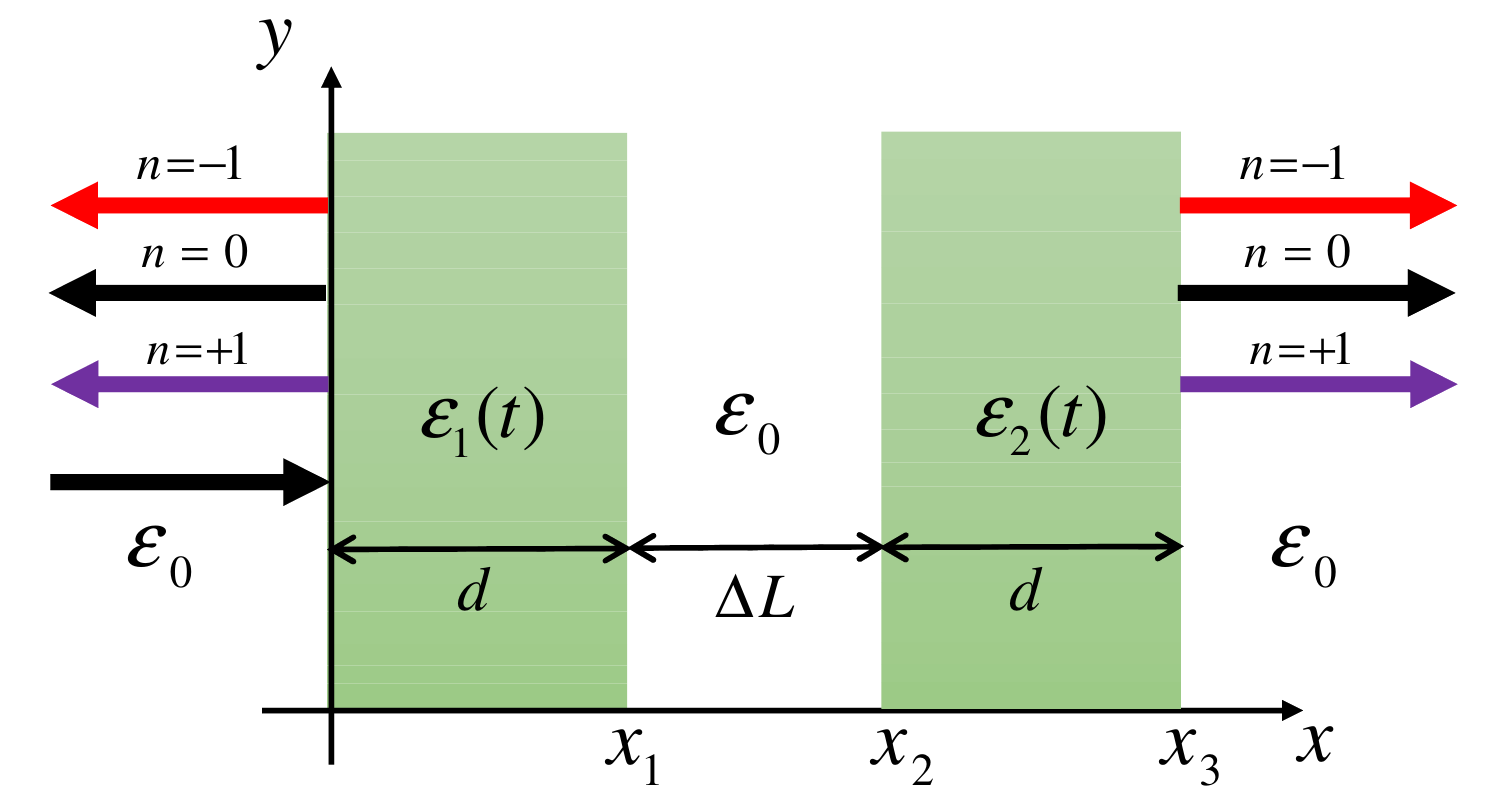}
\caption{Normal incidence on two identical slabs which have a distance from each other.}
\label{fig:two_res}
\end{figure}

In this part, we consider that two FP slabs with the same thickness are placed at a specific distance from each other (Fig. \ref{fig:two_res}). We consider the relative permittivities of both slabs as $\varepsilon_{r,1/2}(t)=\varepsilon_r+\Delta \varepsilon \cos(\Omega t+\varphi_{1/2})$. In the case that $\Delta\varepsilon=0$ (we assume that $\Omega$ is known), based on the distance of these slabs from each other and their thickness, the transmission versus incident frequency can have different plots. Generally speaking, the resonances that can be seen in such medium are due to two factors: the resonances of the slabs, which is theoretically at $\frac{\omega}{\Omega}=\frac{\pi}{\frac{\Omega}{c_0}d\sqrt{\varepsilon_r} }$, and the resonance of the medium between the slabs, which is FP like and theoretically at $\frac{\omega}{\Omega}=\frac{\pi}{\frac{\Omega}{c_0} \Delta L}$.

\begin{figure}[h!]
\centering\includegraphics[width=10cm]{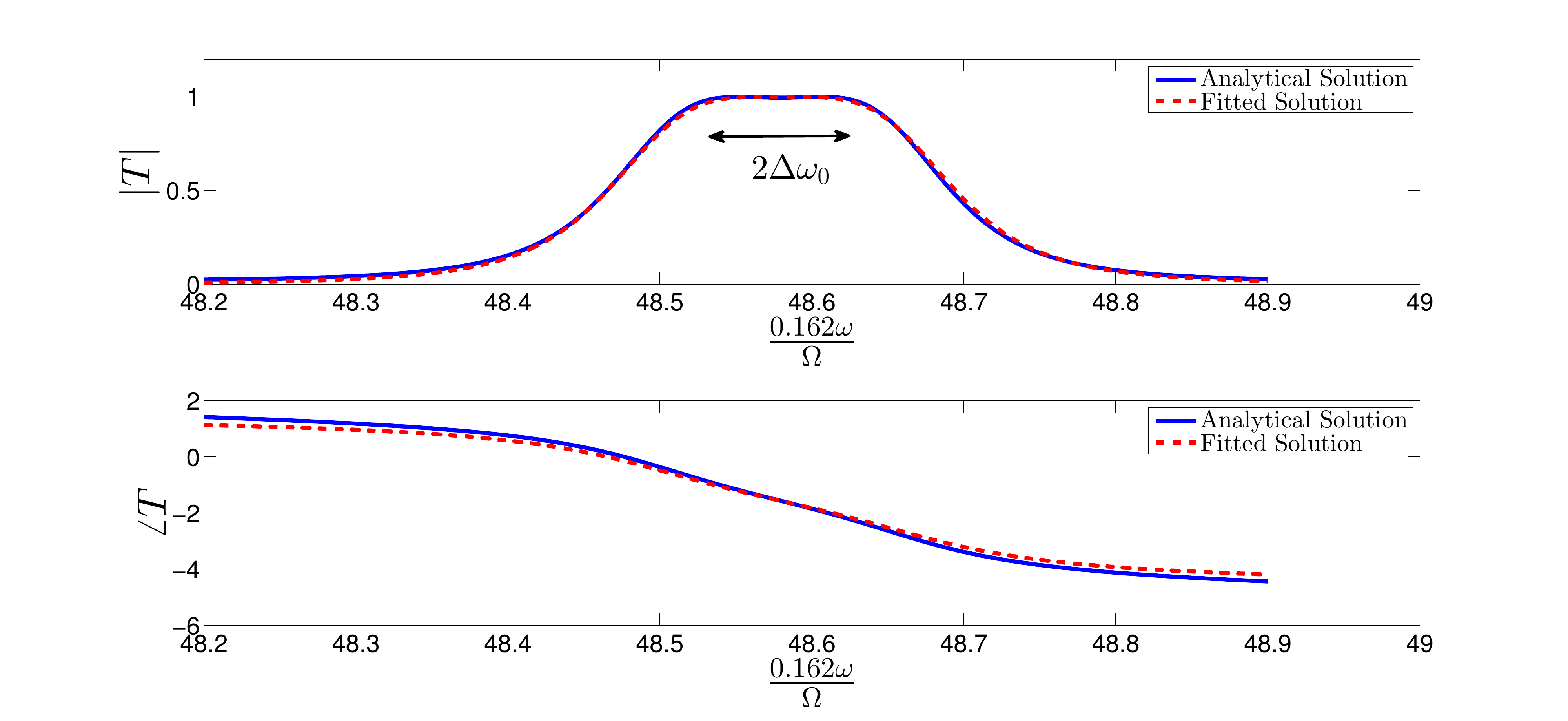}
\caption{Fitted solution for two time-invariant slabs.}
\label{fig:fit_two_no}
\end{figure}

The transmission can be achieved by using TCMT. Here, we choose the thickness of the slabs and the distance between them as $\frac{\Omega}{c_0}d=3.3\times 0.162/\sqrt{\varepsilon_r}$ and $ \frac{\Omega}{c_0}\Delta L=3.3\times 0.162/8$, respectively, and $\varepsilon_r=16$. The transmission of wave from such medium using the analytical method is presented in Fig. \ref{fig:fit_two_no} in solid blue line. 

When two identical resonators are coupled, two modes are formed in the overall system having even and odd symmetries with a red and blue shift from the original resonance frequencies. Our coupled FP slabs also support two orthogonal modes with opposite parity symmetries whose complex frequencies are ${\left( {\frac{0.162\omega }{{\Omega}}} \right)_{mode}} = 48.58 \pm 0.081 - j0.081$ (these frequencies can be determined by using the conventional transfer matrix method \cite{anemogiannis1999determination}). The spatial field distribution of the modes at these frequencies are presented in the Figs. \ref{fig:Mode1} and \ref{fig:Mode2}. 

Using the theory of multi-mode resonators \cite{suh2004temporal}, we can write the TCMT of a two-mode resonator as follows:
\begin{figure}
\centering
\begin{subfigure}[b]{0.6\textwidth}
   \includegraphics[width=1\linewidth]{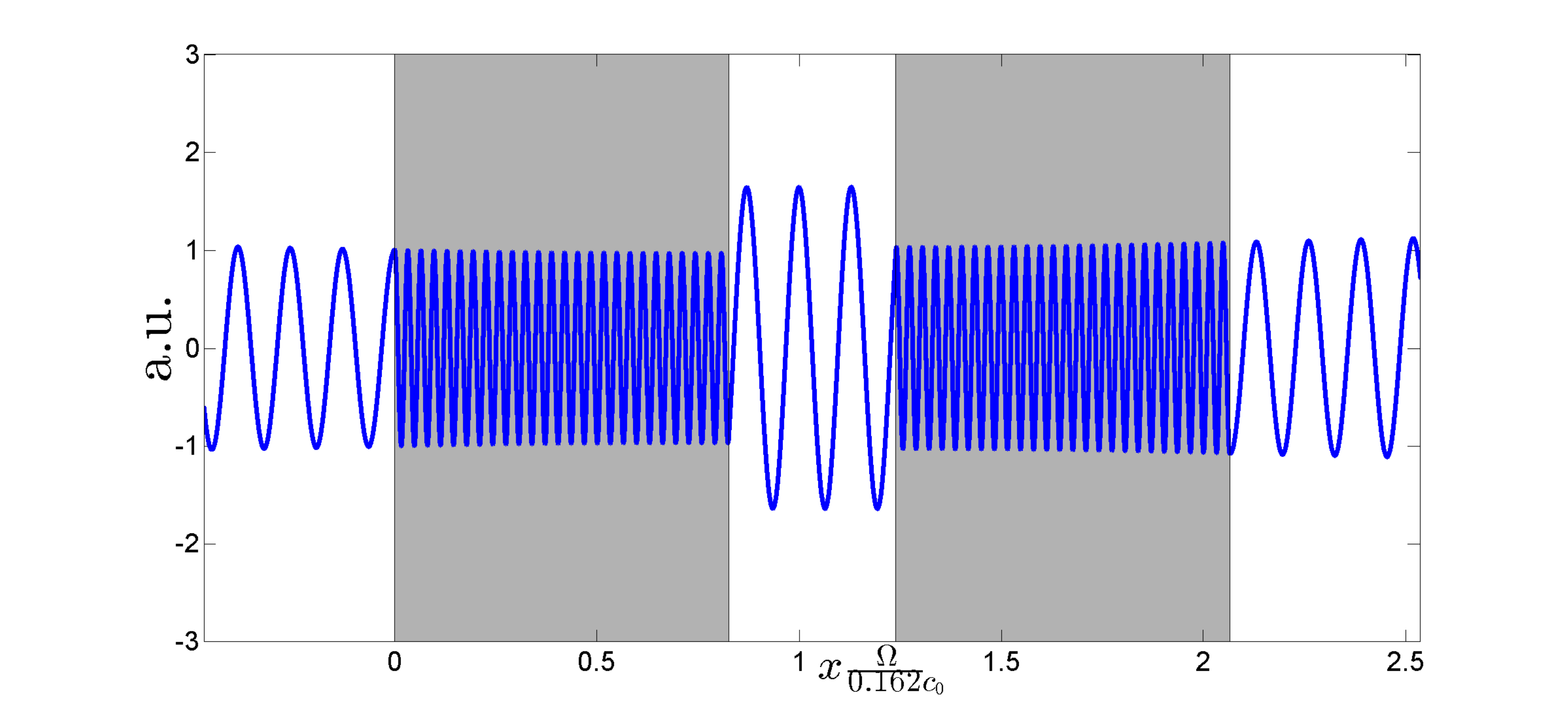}
   \caption{}
   \label{fig:Mode1} 
\end{subfigure}

\begin{subfigure}[b]{0.6\textwidth}
   \includegraphics[width=1\linewidth]{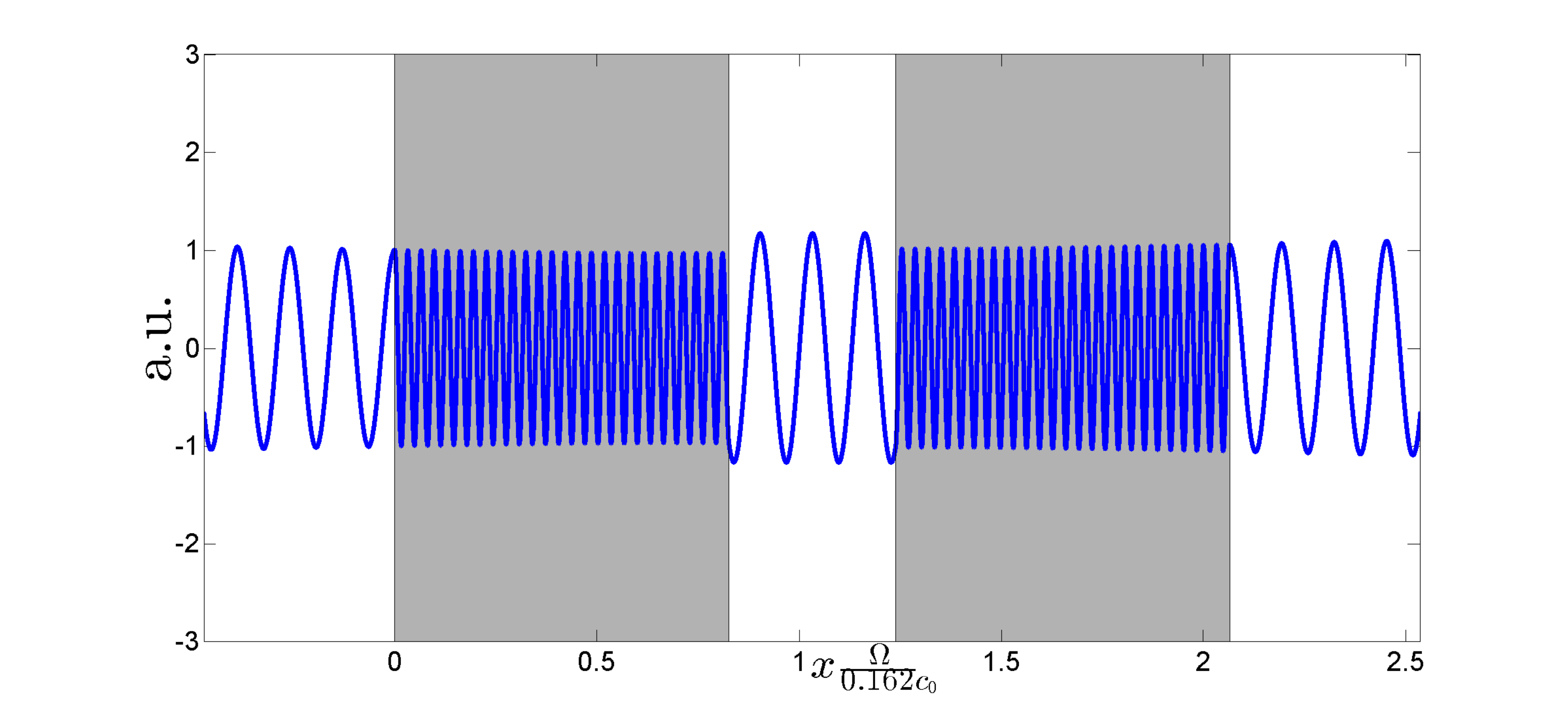}
   \caption{}
   \label{fig:Mode2}
\end{subfigure}
\label{fig:1}
\caption{Field distribution of modes at (a)  ${\left( {\frac{0.162\omega }{{\Omega}}} \right)_{mode}} = 48.58 + 0.081 - j0.081$ (b)  ${\left( {\frac{0.162\omega }{{\Omega}}} \right)_{mode}} = 48.58 - 0.081 - j0.081$.}
\end{figure}

\begin{subequations}
\begin{equation}
\frac{d}{{dt}}{\left| {{a}} \right\rangle _{2 \times 1}} = {(j{\bf{\Omega }} - {\bf{\Gamma }})_{2 \times 2}}\left| {{a}} \right\rangle  + {{\bf{K}}^T}{\left| {{S_ + }} \right\rangle _{2 \times 1}}
\label{eq:28a}
\end{equation}
\begin{equation}
{\left| {{S_ - }} \right\rangle _{2 \times 1}} = {{\bf{C}}_{2 \times 2}}\left| {{S_ + }} \right\rangle  + {{\bf{K}}_{2 \times 2}}\left| {{a}} \right\rangle 
\label{eq:28b}
\end{equation}
\label{eq:28}
\end{subequations}
where $\bf \Omega$ is now a matrix of the resonance frequencies, $\bf \Gamma$ is the matrix of the decaying constants, $\bf K$ is the matrix that couples the incident waves to modes, and $\bf C$ is the matrix that couples incident waves to reflected waves directly. Note that now $\left| {{a}} \right\rangle$ is a $2\times1$ vector containing the energy amplitudes of the two modes, and thus the need for the tensor representation of $\bf \Omega$, $\bf \Gamma$, and $\bf K$. The relations between matrices in this equation are described elaborately in \cite{suh2004temporal}. We have:
\begin{equation}
\begin{aligned}
{{\bf{\Omega }}_{(n,m)}} &= \left\{ {{\omega _0} + ( - 2n + 3)\Delta {\omega _0}} \right\}{\delta _{nm}}\\
{{\bf{\Gamma }}_{(n,m)}} &= \gamma {\delta _{mn}}\\
{\bf{K}} &= j\left[ {\begin{array}{*{20}{c}}
{ + \sqrt \gamma  }&{ + \sqrt \gamma  }\\
{ + \sqrt \gamma  }&{ - \sqrt \gamma  }
\end{array}} \right]
\end{aligned}
\label{eq:29}
\end{equation}
As the resonators are direct coupled, $\bf C$ is the identity matrix. The transmission is therefore:
\begin{equation}
\frac{{S_2^ - }}{{S_1^ + }} = \frac{\gamma }{{j(\omega  - {\omega _0} - \Delta \omega_0 ) + \gamma }} - \frac{\gamma }{{j(\omega  - {\omega _0} + \Delta \omega_0 ) + \gamma }}
\label{eq:30}
\end{equation}
By considering $\omega_0$ as the original resonance frequency of the single slab and $\Delta \omega_0$ as half the splitting frequency (Fig. \ref{fig:fit_two_no}), this function has one fitting parameter, $\gamma$, which can be determined easily. It can be seen that both the analytical and the TCMT method virtually overlap in Fig. \ref{fig:fit_two_no}, where the fitted solution is plotted in dashed red line. 

Having found the TCMT relations of the two direct coupled FP slab resonators for the static structure, i.e. $\Delta \varepsilon=0$, we aim to find the TCMT model for their time-varying version, i.e. $\Delta \varepsilon\neq0$. The procedure is similar to that of the single slab, and the coupling matrix ($\bf K$) and the frequency matrix ($j\mathbf{\Omega}-\mathbf{\Gamma}$) should be perturbed with a time-periodic perturbation (with three terms for each element of the matrices).

\begin{figure}[h!]
\centering\includegraphics[width=10cm]{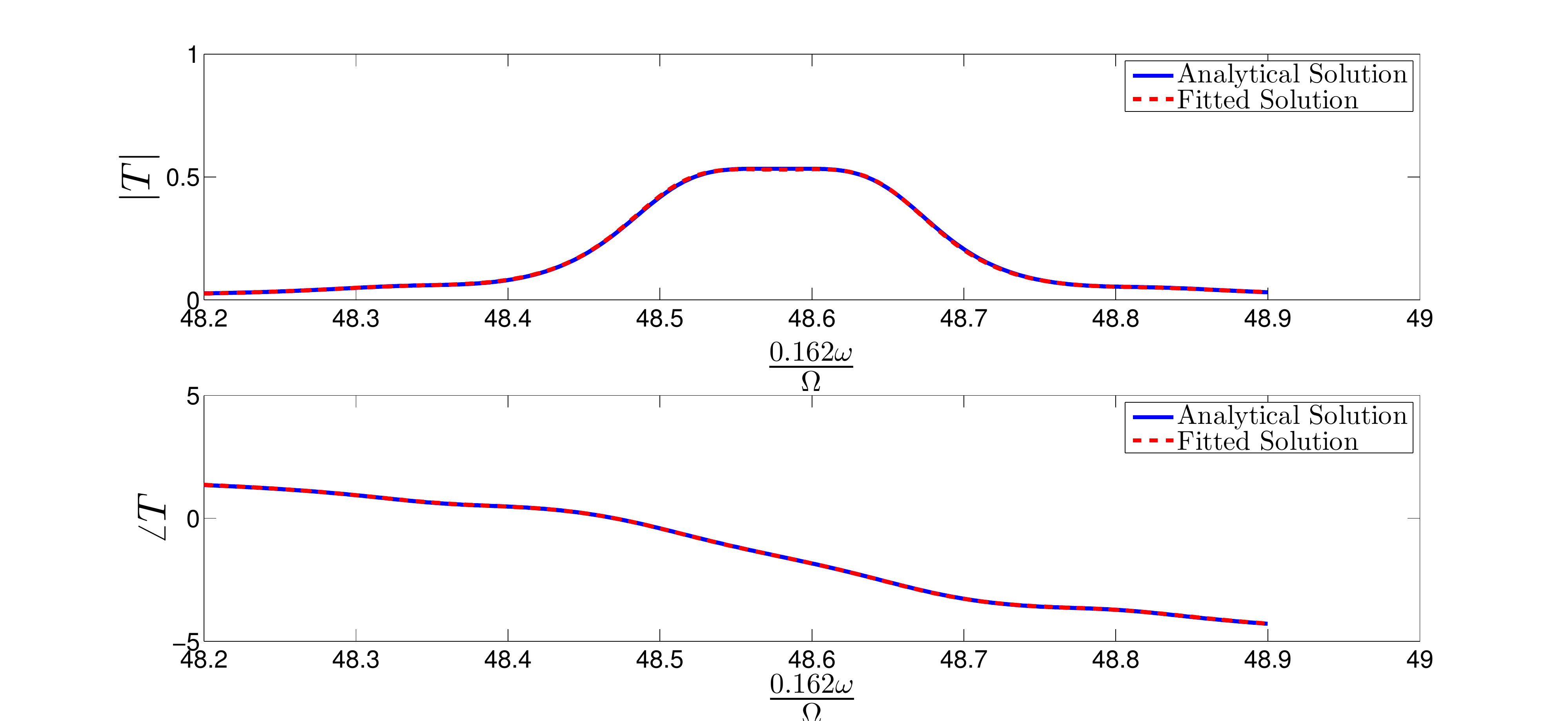}
\caption{Fitted solution for two time-varying slabs, which are sinusoidally modulated with zero phase difference.}
\label{fig:fit_two_mod_0}
\end{figure}

Following similar procedure for perturbing Eq. (\ref{eq:28}), we have
\begin{subequations}
\begin{equation}
\frac{d}{{dt}}\left| {{a}} \right\rangle  = \left( {j{\bf{\Omega }} - {\bf{\Gamma }} + {\bf{W}}(t)} \right)\left| {{a}} \right\rangle  + {\left( {{\bf{K}} + {\bf{D}}(t)} \right)^T}\left| {{S_ + }} \right\rangle 
\label{eq:31a}
\end{equation}
\begin{equation}
\left| {{S_ - }} \right\rangle  = {\bf{C}}\left| {{S_ + }} \right\rangle  + \left( {{\bf{K}} + {\bf{D}}(t)} \right)\left| {{a}} \right\rangle 
\label{eq:31b}
\end{equation}
\label{eq:31}
\end{subequations}
where
\begin{equation}
\begin{aligned}
{\bf{W}}{(t)_{(i,j)}} &= \sum\limits_{n =  - 1}^{ + 1} {\alpha _n^{(i,j)}{e^{ + jn\Omega t}}} \\
{\bf{D}}{(t)_{(i,j)}} &= \sum\limits_{n =  - 1}^{ + 1} {\chi _n^{(i,j)}{e^{ + jn\Omega t}}} \\
{\left| {{a}} \right\rangle _{(i,1)}} &= \sum\limits_{n =  - 1}^{ + 1} {a_n^{(i)}{e^{ + jn(\omega+n\Omega) t}}} \\
\end{aligned}
\label{eq:32}
\end{equation}
Equation (\ref{eq:31a}) can be written in phasor form as:
\begin{equation}
\left( {{{\bf{W}}_0} - \Delta {\bf{W}}} \right)\left( {\left| {{{\bf{a}}_0}} \right\rangle  + \left| {\delta {\bf{a}}} \right\rangle } \right) = \left| {\bf{d}} \right\rangle  + \left| {\delta {\bf{d}}} \right\rangle 
\label{eq:32}
\end{equation}
where
\begin{equation}
\begin{aligned}
{{\bf{W}}_0} &= \left[ {\begin{array}{*{20}{c}}
{{\bf{W}}_{^{3 \times 3}}^{(1,1)}}&{{{\bf{0}}_{3 \times 3}}}\\
{{{\bf{0}}_{3 \times 3}}}&{{\bf{W}}_{^{3 \times 3}}^{(2,2)}}
\end{array}} \right],{\bf{W}}_{^{(n,m)}}^{(1,1)/(2,2)} = \left\{ {j\left( {\omega  - {\omega _0} \mp \Delta {\omega _0} + (2 - n)\Omega } \right) + \gamma } \right\}{\delta _{nm}}\\
\Delta {\bf{W}} &= \left[ {\begin{array}{*{20}{c}}
{\Delta {\bf{W}}_{^{3 \times 3}}^{(1,1)}}&{\Delta {\bf{W}}_{^{3 \times 3}}^{(1,2)}}\\
{\Delta {\bf{W}}_{^{3 \times 3}}^{(2,1)}}&{\Delta {\bf{W}}_{^{3 \times 3}}^{(2,2)}}
\end{array}} \right],\Delta {\bf{W}}_{^{(n,m)}}^{(i,j)} = \alpha _{m - n}^{(i,j)}\\
\left| {\bf{d}} \right\rangle  &= \left[ {0,{{\bf{K}}_{(1,1)}},0,0,{{\bf{K}}_{(1,2)}},0} \right]\\
\left| {\delta {\bf{d}}} \right\rangle  &= \left[ {{\chi _{ + 1}}^{(1,1)},{\chi _0}^{(1,1)},{\chi _{ - 1}}^{(1,1)},{\chi _{ + 1}}^{(1,2)},{\chi _1}^{(1,2)},{\chi _{ - 1}}^{(1,2)}} \right]
\end{aligned}
\label{eq:32}
\end{equation}
In writing this equation, again we divided $\left| {\bf{a}} \right\rangle$ into two terms to avoid inversing full matrices. The unperturbed solution can be written as $\left| {\bf{a}}_0 \right\rangle={{\bf{W}}_0}^{ - 1}\left| {\bf{d}} \right\rangle $. Once more, we ignore the term $\Delta {\bf{W}}{\left| {\delta {\bf{a}}} \right\rangle }$ to find an approximate solution for ${\left| {\delta {\bf{a}}} \right\rangle }$. By substituting  ${\left| { {\bf{a}}} \right\rangle= \left| { {\bf{a}_0}} \right\rangle+\left| { {\delta\bf{a}}} \right\rangle}$ into Eq. (\ref{eq:31b}) and performing some algebraic manipulations, the transmission at the incidence frequency, the $-1$ harmonic, and the $+1$ harmonic can be found respectively as:
\begin{subequations}
\begin{equation}
{\left. {\frac{{S_2^ - }}{{S_1^ + }}} \right|_\omega } = {T_0} + \sum\limits_{n =  - 1}^{ + 1} {\left\{ {\frac{{{z_n}}}{{j\left( {\omega  - {\omega _0} - \Delta {\omega _0} + n\Omega } \right) + \gamma }} + \frac{{{w_n}}}{{j\left( {\omega  - {\omega _0} + \Delta {\omega _0} + n\Omega } \right) + \gamma }}} \right\}} 
\label{eq:33a}
\end{equation}
\begin{equation}
{\left. {\frac{{S_2^ - }}{{S_1^ + }}} \right|_{\omega  - \Omega }} = \sum\limits_{n =  - 1}^0 {\left\{ {\frac{{z_{_n}^{( - 1)}}}{{j\left( {\omega  - {\omega _0} - \Delta {\omega _0} + n\Omega } \right) + \gamma }} + \frac{{w_{_n}^{( - 1)}}}{{j\left( {\omega  - {\omega _0} + \Delta {\omega _0} + n\Omega } \right) + \gamma }}} \right\}} 
\label{eq:33b}
\end{equation}
\begin{equation}
{\left. {\frac{{S_2^ - }}{{S_1^ + }}} \right|_{\omega  + \Omega }} = \sum\limits_{n = 0}^{ + 1} {\left\{ {\frac{{z_{_n}^{( + 1)}}}{{j\left( {\omega  - {\omega _0} - \Delta {\omega _0} + n\Omega } \right) + \gamma }} + \frac{{w_{_n}^{( + 1)}}}{{j\left( {\omega  - {\omega _0} + \Delta {\omega _0} + n\Omega } \right) + \gamma }}} \right\}} 
\label{eq:33c}
\end{equation}
\label{eq:33}
\end{subequations}
where $z_n$ and $w_n$ are unknown coefficients to be fitted, and $T_0$ is the transmission which was determined in Eq. (\ref{eq:30}). It is worth noting that we have simplified the final expression in Eq. (\ref{eq:33}). In fact, each of the $z_n$ and $w_n$ consists of many perturbation terms; however, we consider their total contribution as a single complex number. These unknown parameters can be calculated from the transmission response obtained from the analytical method at the frequencies of the denominators of Eq. (\ref{eq:33}). These values provide six complex numbers, which are sufficient for finding the six unknown complex parameters. For the previous example, we first consider the permittivty of slabs as $\varepsilon_r=16$, $\Delta\varepsilon_r=0.075$, and $\varphi_1=\varphi_2=0$ and a modulation frequency which is equal to the splitting frequency, i.e. $\Omega=2\Delta \omega_0$. The transmission at the incident frequency is plotted in Fig. \ref{fig:fit_two_mod_0}, at which the accuracy of the result is acceptable. It should also be noted that for the analytical method we have considered sufficient number of harmonics, so the result has converged. It can be seen here that although both slabs are time-varying, one does not achieve non-reciprocity due to the spatial symmetry of the structure. In what follows, we demonstrate an interesting case of non-reciprocity with a similar structure.

\section{Non-reciprocity with Two Cascaded Time-Varying Quadrature Phase Fabry-Perot Slabs}
It is shown in many articles that a spatio-temporal modulated permittivity can provide non-reciprocity at the incident frequency \cite{taravati2019generalized,taravati2019space,chamanara2019simultaneous,yu2009complete,shaltout2015time,nagulu2018nonreciprocal,taravati2017nonreciprocal2,taravati2017nonreciprocal,wang2012non,wang2014time},  meaning that the incident wave can pass the structure from one side without losing any power in the central frequency, but it will be blocked when it is incident from the other side. Nevertheless, this method of modulation is hard to be implemented because of partial reflections of the modulating wave at boundaries, so a pure traveling wave modulation cannot be obtained. Here, it will be shown that non-reciprocity can be achieved by using two time-varying FP slabs, where neither of them has any spatial modulation, and only exhibits temporal modulation.

\begin{figure}[h!]
\centering\includegraphics[width=10cm]{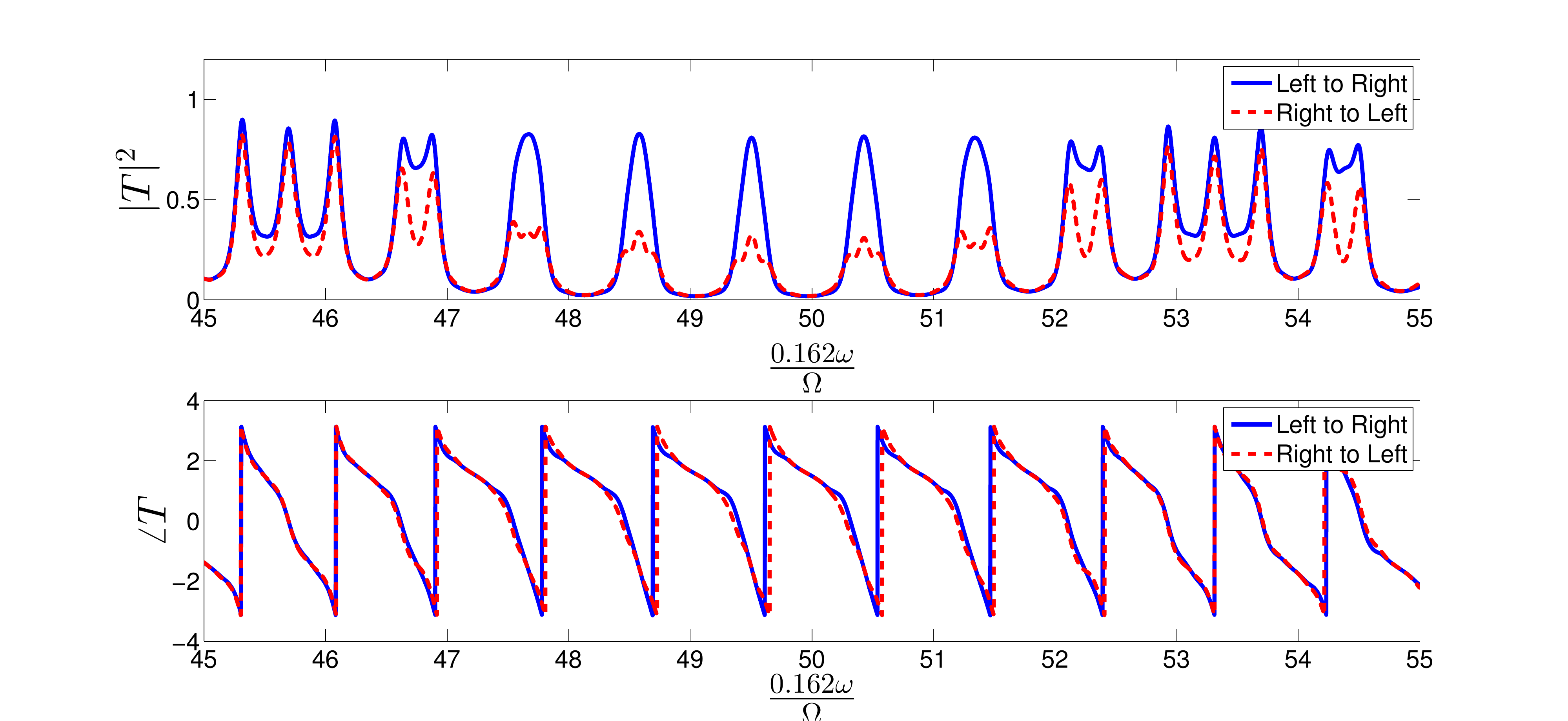}
\caption{Fitted solution for two time-varying slabs, which are sinusoidally modulated with $\pi/2$ phase difference.}
\label{fig:Mod_Res_Res}
\end{figure}

In our previous example of two slabs, we can achieve non-reciprocity by allowing a phase difference between the permittivity functions of the two slabs, i.e. $\varphi_1\neq \varphi_2$. Based on our observations, the best choice for this phase difference is quadrature phase, i.e. $\varphi_1=0$ and $\varphi_2=\pi/2$. Again, let us consider the modulation frequency to be equal to the splitting frequency, i.e. $\Omega = 2\Delta \omega_0$. The transmission versus incidence frequency from right and left can be seen in Fig. \ref{fig:Mod_Res_Res}, where non-reciprocity is now clearly evident even at the incidence frequency (zeroth harmonic). 

Based on the perturbed TCMT which was derived in the previous section (Eq. (\ref{eq:33})), a reason for such non-reciprocity can be provided. By using the stated procedure for fitting, the fitting parameters are obtained as in Table \ref{table:1}.
\begin{table}\centering
\caption{Fitting parameters}
\ra{1.3}
\begin{tabular}{cccc}
\toprule
 &Left to Right  &\phantom{ab} &Right to Left\\
\midrule 
$w_0$      &$\sim 0.01\angle{-154^\circ}$    && $\sim0.053\angle{-150^\circ }$     \\
$z_0$          &$\sim0.01\angle{-26^\circ}$         && $\sim0.053\angle{-30^\circ}$       \\
$w_1$       & $\sim0.0012\angle{140^\circ}$     &&$\sim 0$  \\
$z_1$       &  $\sim0.0012\angle{40^\circ}$     &&$\sim 0$  \\
\bottomrule
\end{tabular}
\label{table:1}
\end{table}

\begin{figure}[h!]
\centering\includegraphics[width=10cm]{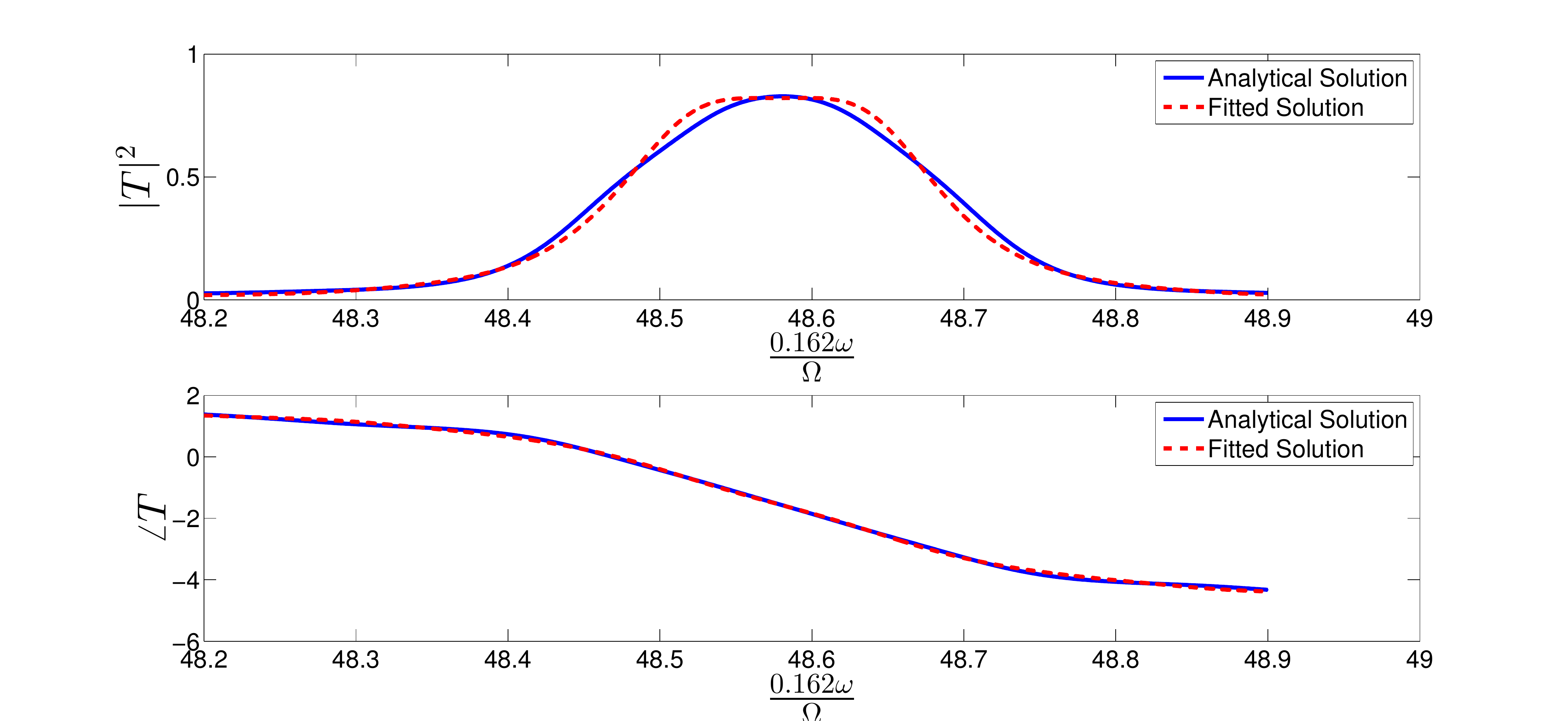}
\caption{Fitted solution for two time-varying slabs, which are sinusoidally modulated with $\pi/2$ phase difference, left to right transmission.}
\label{fig:L_to_R}
\end{figure}

\begin{figure}[h!]
\centering\includegraphics[width=10cm]{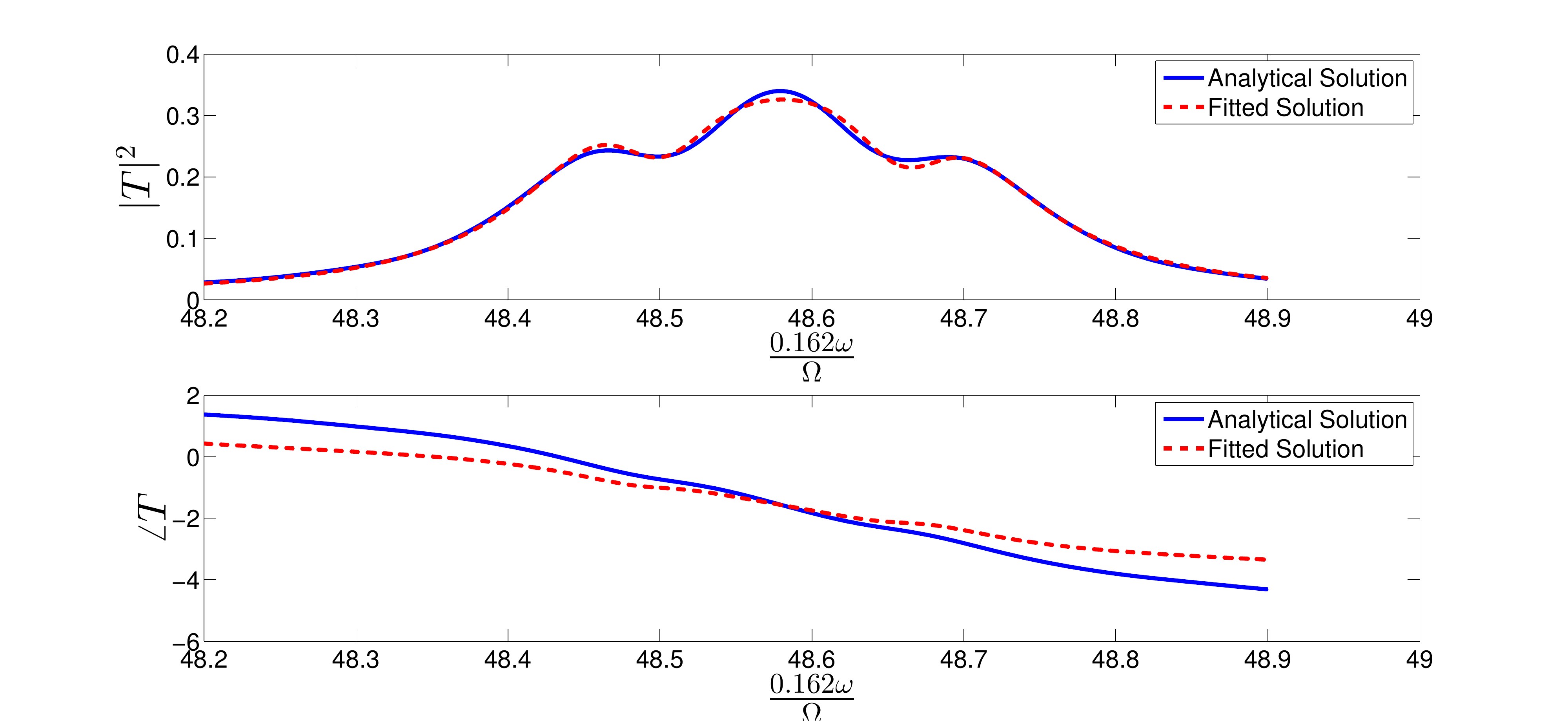}
\caption{Fitted solution for two time-varying slabs, which are sinusoidally modulated with $\pi/2$ phase difference, right to left transmission.}
\label{fig:R_to_L}
\end{figure}

The fitted solution for transmission from left to right and right to left can be seen in the Figs. \ref{fig:L_to_R} and \ref{fig:R_to_L}, and an acceptable overlap can be observed between the two methods. As it can be seen from the Table \ref{table:1}, we have $w_n=-z_n^*$ . Furthermore, it can be observed that $|w_1|\simeq |z_1| \simeq 0$, which can be expected, because the time varying part of the permittivity has a very small amplitude in comparison to the background permittivity, i.e. $0.075/16\simeq 0.5\%$; as a result, higher orders of perturbation terms have negligible levels. Nevertheless, there is a great difference between $z_0,w_0$ for left to right and right to left transmission, which can account for the difference in their transmission. In other words, from left to right transmission ($\varphi=0\to \pi/2$ ) the perturbation terms do not play an important role, while they become significant for the right to the left transmission ($\varphi=\pi/2 \to 0$ ).

\begin{figure}[h!]
\centering\includegraphics[width=10cm]{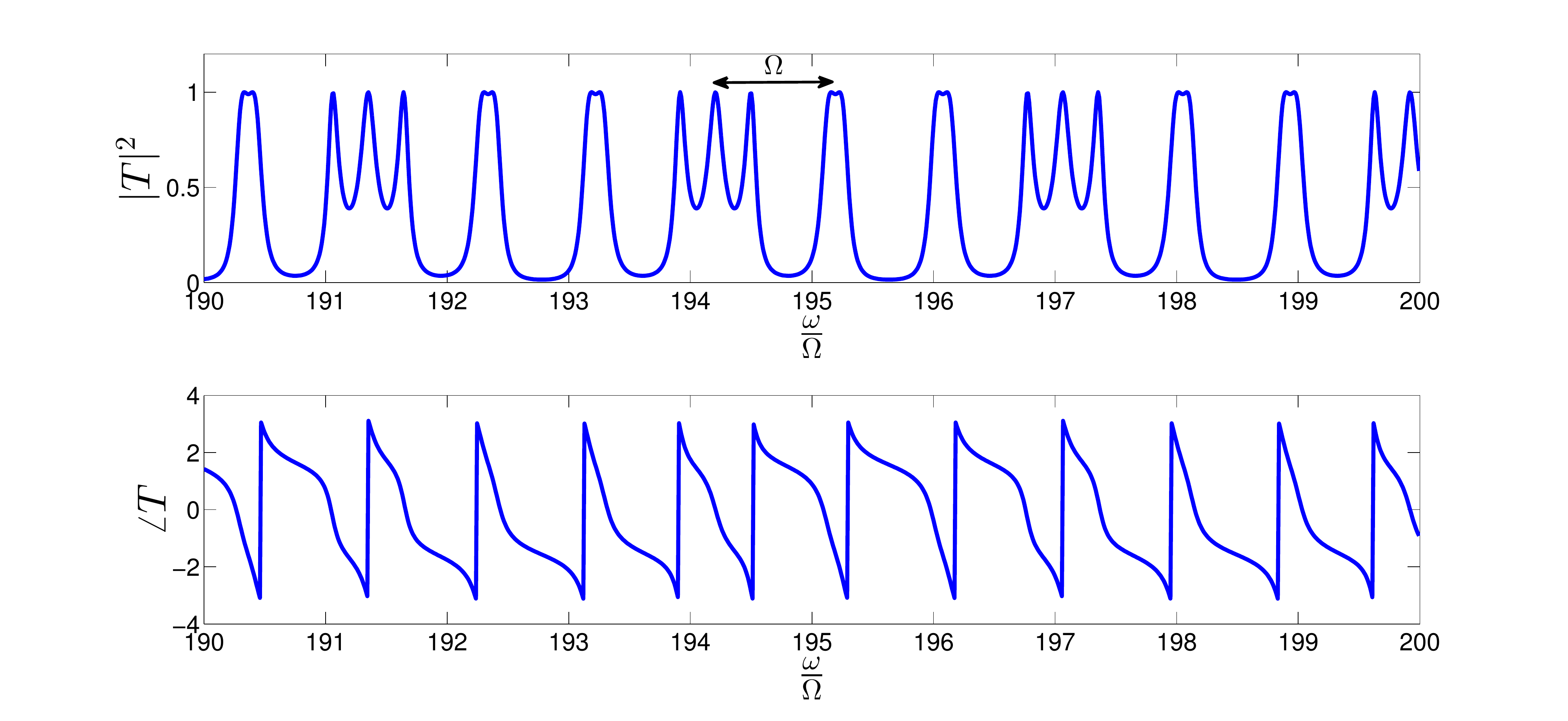}
\caption{Transmission from two time-invariant slabs with $\frac{\Omega}{c_0}d=3.3/\sqrt{16}$ and $\frac{\Omega}{c_0}\Delta L=3.3/3$.}
\label{fig:great_non_rec_no}
\end{figure}

We can achieve significant non-reciprocity with our simple structure at hand, i.e. transmission of approximately $1$ from one side to another and approximately $0$ for the reverse direction in the incident frequency. For two slabs with $\Delta \varepsilon=0$ ($\Omega$ is again known), $\frac{\Omega}{c_0}d=3.3/\sqrt{16}$, and $\frac{\Omega}{c_0}\Delta L=3.3/3$, the transmission can be seen in Fig. \ref{fig:great_non_rec_no} (obtained using the analytical method). In this figure, three adjacent resonances can be seen, which are due to the fact that the resonance frequency of the medium between two slabs coincides with the resonance frequency of the FP slabs, i.e. $m\frac{\pi }{{d\sqrt {{\varepsilon _r}} }} = n\frac{\pi }{{\Delta L}}$ for $\left\{ {m,n} \right\} \in \mathbb{N}$.

\begin{figure}[h!]
\centering\includegraphics[width=10cm]{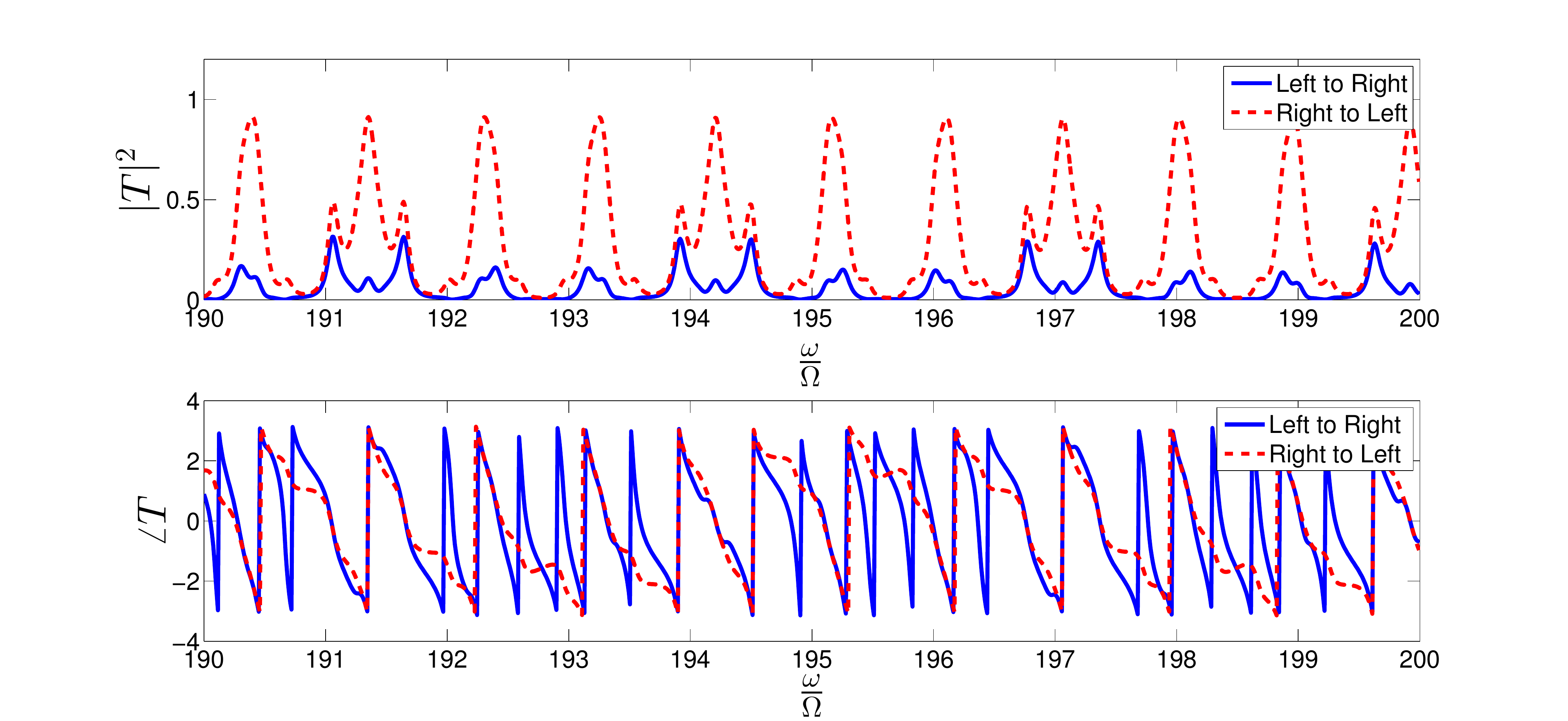}
\caption{Transmission from two time-varying slabs with $\frac{\Omega}{c_0}d=3.3/\sqrt{16}$ and $\frac{\Omega}{c_0}\Delta L=3.3/3$, which have a $\pi/2$ phase difference.}
\label{fig:great_non_rec}
\end{figure}

Now, for achieving a noticeable non-reciprocity, we consider the modulation frequency to be what is shown in the Fig. \ref{fig:great_non_rec_no}. By doing so, and having a $\pi/2$ phase difference between two time varying parts of left and right slabs (their amplitude is $0.075$), the transmission from the right to the left and vice versa can be seen in Fig. \ref{fig:great_non_rec}, where a noticeable non-reciprocity is observed. A TCMT fitted solution can also be provided; however, it would require developing the formulation for a 5-mode resonator, which makes the solution very complicated. 

\begin{figure}[!tbp]
  \begin{subfigure}[b]{0.5\textwidth}
    \includegraphics[width=\textwidth]{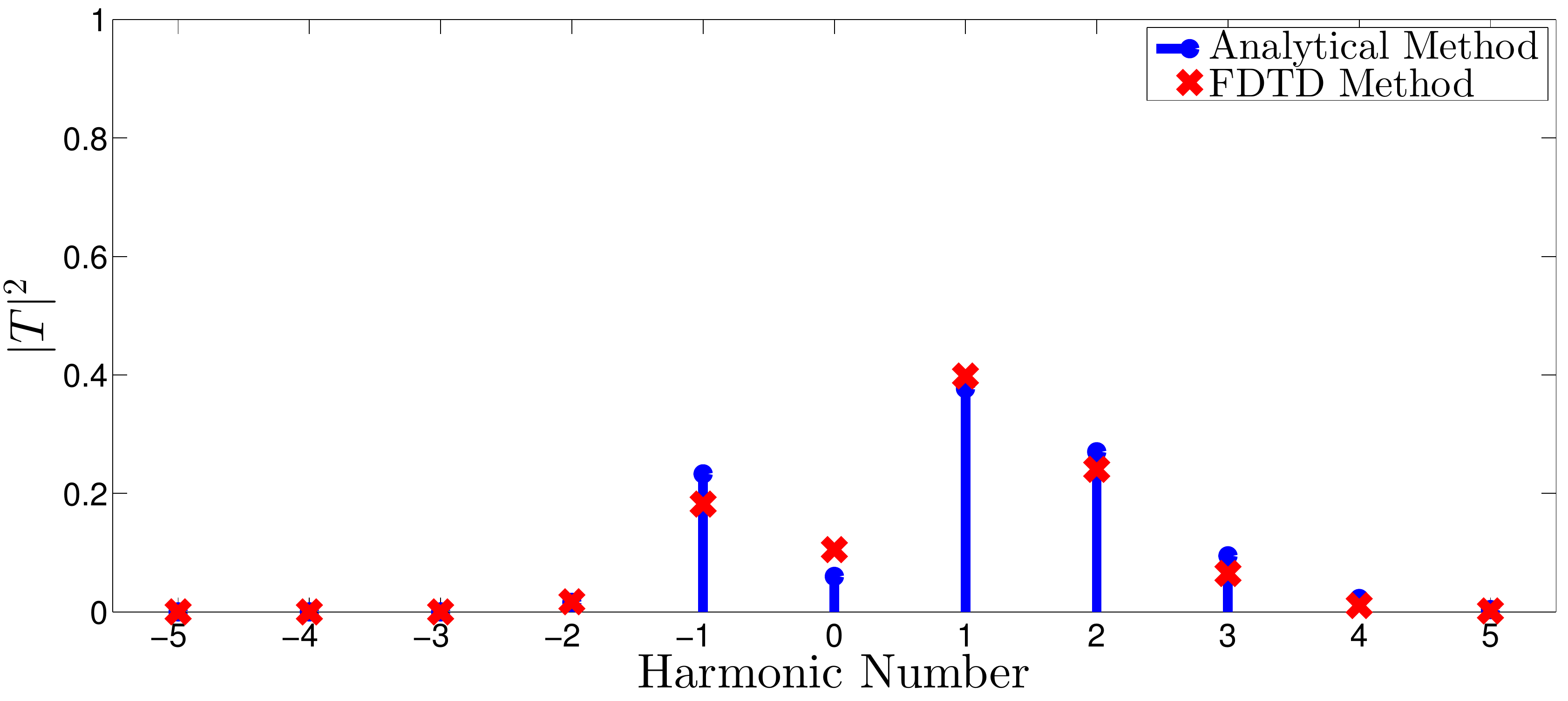}
	\caption{}
    \label{fig:left_right}
  \end{subfigure}
  \hfill
  \begin{subfigure}[b]{0.5\textwidth}
    \includegraphics[width=\textwidth]{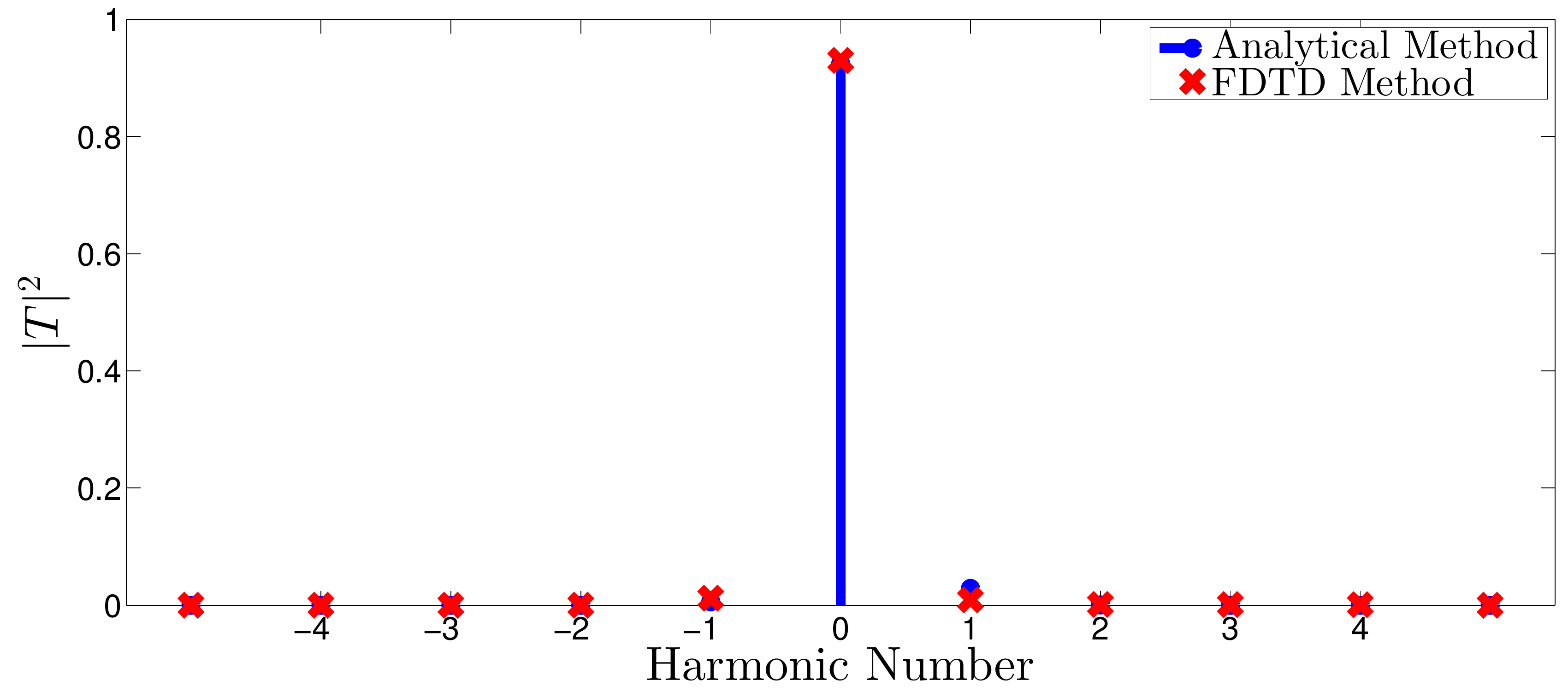}
    \caption{}
    \label{fig:right_left}
  \end{subfigure}
  \caption{(a) Left to right and (b) right to left transmission from two time-varying slabs with $\frac{\Omega}{c_0}d=3.3/\sqrt{16}$, $\frac{\Omega}{c_0}\Delta L=3.3/3$, and modulation amplitudes equal to 4. Analytical solution is in solid blue line, and FDTD solution is in red squares.}
\end{figure}

To verify that the observed non-reciprocity is indeed valid, we developed an in-house FDTD code for validation. Consider the previous example with a higher modulation amplitude ($0.075$ changed to $4$) in order to observe non-reciprocity in lower frequencies (and thus reduce FDTD simulation time). The left to right and right to left transmission versus harmonic numbers at $\frac{\omega}{\Omega}=3.86$ can be seen in the Figs. \ref{fig:left_right} and \ref{fig:right_left}, respectively, where the analytical solution is in solid blue line, and the FDTD solution is in red squares. The considerable overlap of the numerical results and the analytical results validates the presence of large non-reciprocity in such scenario. 

\section{Conclusion}
It was demonstrated that significant non-reciprocity can be achieved by using only two equal-sized FP slab resonators whose permittivity functions only have a sinusoidal temporal dependency, but with a quadrature phase difference, and our results were validated using in-house FDTD code. In our proposed method, the time varying media do not require spatial modulation, i.e. propagating-wave modulations of permittivity are not used, which is a great advantage over existing methods. To find the forward and backward wave amplitudes (at all harmonics) due to a wave incident on multiple time-varying slabs, a generalized transfer matrix method was proposed, which can be used for normal incidence. In addition, to have a simple model of the medium (for finding transmission and reflection), a time-perturbed temporal coupled mode theory is developed, which can explain the behavior of a single time-varying slab, as well as time-perturbed formulation of multi-mode resonators is proposed to explain the behavior of multiple time-varying slabs. It is shown that both the analytical and the latter method virtually overlap.

\section*{Appendix A: Generalized Transfer Matrix Method}
\newcommand{\hbAppendixPrefix}{A}
\renewcommand{\thefigure}{\hbAppendixPrefix\arabic{figure}}
\setcounter{figure}{0}
\renewcommand{\thetable}{\hbAppendixPrefix\arabic{table}} 
\setcounter{table}{0}
\renewcommand{\theequation}{\hbAppendixPrefix\arabic{equation}} 
\setcounter{equation}{0}

\label{section:app}
Here, we try to find a general solution for wave propagation in a 1D medium with multiple slabs whose permittivities have arbitrary periodic functions of time with the same modulation frequency (Fig. \ref{fig:structure}). By starting from Maxwell's Equations in a 1D medium, we can write:
\begin{subequations}
\begin{equation}
\frac{{\partial {H_y}}}{{\partial x}}  = \frac{{\partial {D_z}}}{{\partial t}}\\
\label{eq:1a}
\end{equation}
\begin{equation}
\frac{{\partial {E_z}}}{{\partial x}} = \mu_0\frac{{\partial {H_y}}}{{\partial t}}\\
\label{eq:1b}
\end{equation}
\label{eq:1}
\end{subequations}
By using Floquet-Bloch theorem in time, we can write ${E_z}(x,t) = \sum\limits_n {{E_{z,n}}(x){e^{j(\omega  + n\Omega )t}}}  + c.c.$, whereat $\omega$ is the frequency of the incident wave. The same method can be used for $H_{y}(x,t)$. By inserting the fields in the Eq. (\ref{eq:1}) and doing some algebraic manipulations, one can write:
\begin{subequations}
\begin{equation}
\frac{{\partial {H_{y,n}}(x)}}{{\partial x}} =  j{\varepsilon _0}\sum\limits_m {\varepsilon _{(n - m)}}(x)(\omega  +n\Omega ){E_{z,m}}(x)
\label{eq:3b}
\end{equation}
\begin{equation}
\frac{{\partial {E_{z,n}}(x)}}{{\partial x}} = j{\mu _0}(\omega  + n\Omega ){H_{y,n}}(x)
\label{eq:3c}
\end{equation}
\label{eq:3}
\end{subequations}
where we have considered a Fourier series (in time) for relative permittivity in the form of ${\varepsilon _r}(x,t) = \sum\limits_n {{\varepsilon _n}(x){e^{jn\Omega t}}} $. This equation can be written in each section, Because the permittivity does not change with $x$ for $x_i<x<x_{i+1}$. By introducing the following variables:
\begin{equation}
\begin{aligned}
{{\bf{E}}_N} &= {\left( {{E_{z,N}},...,{E_{z,0}},...,{E_{z, - N}}} \right)^T}\\
{{\bf{H}}_N} &= {\left( {{H_{y,N}},...,{H_{y,0}},...,{H_{y, - N}}} \right)^T}
\end{aligned}
\label{eq:4}
\end{equation}
the Eq. (\ref{eq:3}) can be written in the following form:
\begin{equation}
\frac{\partial }{{\partial \xi(x) }}\left[ \begin{array}{l}
{{\bf{E}}_N}\\
{{\bf{H}}_N}
\end{array} \right]_i = \mathbf{W}_i \left[ \begin{array}{l}
{{\bf{E}}_N}\\
{{\bf{H}}_N}
\end{array} \right]_i=\left[ {\begin{array}{*{20}{c}}
\emptyset &{{{\bf{W}}_{E,i}}}\\
{{{\bf{W}}_{H,i}}}&\emptyset 
\end{array}} \right]\left[ \begin{array}{l}
{{\bf{E}}_N}\\
{{\bf{H}}_N}
\end{array} \right]_i
\label{eq:5}
\end{equation}
where the subscript $i$ indicates that the fields are related to the  $x_i<x<x_{i+1}$ and 
\begin{equation}
\begin{aligned}
\xi (x) &= \frac{\Omega }{{{c_0}}}x\\
\mathbf{W}_{E,i,(n,m)} &= j{\eta _0}(\frac{\omega }{\Omega } + n){\delta _{nm}}\\
{{\bf{W}}_{H,i,(n,m)}} &= \frac{j}{{{\eta _0}}}{\varepsilon _{i,(n - m)}}(\frac{\omega }{\Omega } + n)
\end{aligned}
\label{eq:6}
\end{equation}
Now, because $\mathbf{W}_i$ is independent of $x$ for $x_i<x<x_{i+1}$, the Eq. (\ref{eq:5}) has a solution based on its eigenvalues and eigenvectors:
\begin{equation}
{\left[ {\begin{array}{*{20}{l}}
{{{\bf{E}}_N}}\\
{{{\bf{H}}_N}}
\end{array}} \right]_i} = \sum\limits_{m =  - N}^{ + N} {c_{m,i}^\pm {\bf{V}}_{m,i}^ \pm {e^{ \mp {\lambda _{m,i}}(\xi (x) - \xi ({x_i}))}}}  
\label{eq:7}
\end{equation}
where $\mathbf{V}_{m,i}$, $\lambda _{m,i}$, and $c_{m,i}$ are the $m$th eigenvector, eigenvalue, and unknown coefficient, respectively, which are related to $x_i<x<x_{i+1}$. Because of the special form of the matrix $\mathbf{W}$, if $\lambda _{m,i}$ be an eigenvalue, $-\lambda _{m,i}$ is an eigenvalue as well (the proof is straightforward). In addition, $\pm$ signs correspond to a backward/backward wave, and eigenvalues are pure imaginary because of the normal incidence. Now, to determine unknown coefficients, continuity of transverse fields at each boundary ($x=x_{i,i+1}$) can be used for all sections. By doing some algebraic manipulations, one can easily receive in: 
\begin{equation}
\begin{array}{l}
\left[ {\begin{array}{*{20}{c}}
{\bf{T}}\\
{\bf{0}}
\end{array}} \right] = \left\{ {\prod\limits_{i = 0}^M {{{\bf{Q}}_{(M - i) \to (M + 1 - i)}}} } \right\}\left[ {\begin{array}{*{20}{c}}
{{\bf{C}}_{(0)}^ + }\\
{\bf{R}}
\end{array}} \right]\\
\left[ {\begin{array}{*{20}{c}}
{{\bf{C}}_{(i + 1)}^ + }\\
{{\bf{C}}_{(i + 1)}^ - }
\end{array}} \right] = {{\bf{Q}}_{i \to i + 1}}\left[ {\begin{array}{*{20}{c}}
{{\bf{C}}_{(i)}^ + }\\
{{\bf{C}}_{(i)}^ - }
\end{array}} \right]\\
{{\bf{Q}}_{i \to i + 1}} = {\left[ {{\bf{V}}_{(i + 1)}^ + |{\bf{V}}_{(i + 1)}^ - } \right]^{ - 1}}\left[ {{\bf{V}}_{(i)}^ + |{\bf{V}}_{(i)}^ - } \right]\left[ {\begin{array}{*{20}{c}}
{{{\bf{X}}_{(i)}}}&0\\
0&{{\bf{X}}_{_{(i)}}^{ - 1}}
\end{array}} \right]
\end{array}
\label{eq:100}
\end{equation}
where ${{\bf{Q}}_{i \to i + 1}}$ is the generalized transfer matrix, ${{\bf{C}}_{(i)}^ \pm }$ is the column vector of forward/backward waves in the $i$th section, and 
\begin{equation}
\begin{aligned}
{{\bf{X}}_{(i),(m,n)}}{\rm{ }} &= {e^{ - {\lambda _{(i),(N + 1 - n)}}{d_{(i)}}}}{\delta _{m,n}}\\
{\bf{V}}_{(i)}^ \pm  &= \left[ {{\bf{V}}_{ + N,(i)}^ \pm ,...,{\bf{V}}_{0,(i)}^ \pm ,...,{\bf{V}}_{ - N,(i)}^ \pm } \right]\\
\end{aligned}
\end{equation}

\bibliographystyle{ieeetr}
\bibliography{sample}

\end{document}